\documentclass[pre,twocolumn,preprintnumbers,amsmath,amssymb,nopacs,nofootinbib,floatfix]{revtex4}

\usepackage{graphicx, bm, xcolor, physics, hyperref}
\usepackage[normalem]{ulem}

\makeatletter

\def\graphicscale{\twocolumn@sw{0.3}{0.4}}
\def\graphicthreescale{\twocolumn@sw{0.3}{0.4}}

\begin{document}

\title{Quantum many-body spin rings coupled to ancillary spins: \\
  The sunburst quantum Ising model}

\author{Alessio Franchi}
\affiliation{Dipartimento di Fisica dell'Universit\`a di Pisa
        and INFN, Largo Pontecorvo 3, I-56127 Pisa, Italy}

\author{Davide Rossini}
\affiliation{Dipartimento di Fisica dell'Universit\`a di Pisa
        and INFN, Largo Pontecorvo 3, I-56127 Pisa, Italy}

\author{Ettore Vicari}
\altaffiliation{Authors are listed in alphabetic order.}
\affiliation{Dipartimento di Fisica dell'Universit\`a di Pisa
        and INFN, Largo Pontecorvo 3, I-56127 Pisa, Italy}

\date{\today}

\begin{abstract}
  We study the ground-state properties of a quantum ``sunburst
  model'', composed of a quantum Ising spin-ring in a transverse
  field, symmetrically coupled to a set of ancillary isolated qubits,
  to maintain a residual translation invariance and also a
  $\mathbb{Z}_2$ symmetry.  The large-size limit is taken in two
  different ways: either by keeping the distance between any two
  neighboring ancillary qubits fixed, or by fixing their number while
  increasing the ring size.  Substantially different regimes emerge,
  depending on the various Hamiltonian parameters: for small energy
  scale $\delta$ of the ancillary subsystem and small ring-qubits
  interaction $\kappa$, we observe rapid and nonanalytic changes in
  proximity of the quantum transitions of the Ising ring, both
  first-order and continuous, which can be carefully controlled by
  exploiting renormalization-group and finite-size scaling frameworks.
  Smoother behaviors are instead observed when keeping $\delta>0$
  fixed and in the Ising disordered phase. The effect of an increasing
  number $n$ of ancillary spins turns out to scale proportionally to
  $\sqrt{n}$ for sufficiently large values of $n$.
\end{abstract}

\maketitle

% ========================= BODY =========================

\section{Introduction}
\label{intro}

The recent amazing progress on the control of physical systems at the
nano scale has paved the way toward deep investigations of 
quantum properties of matter and the more complex problem
of monitoring the coherent quantum dynamics of mutually coupled
systems, addressing
energy interchanges and the relative decoherence properties among the
various subsystems~\cite{Zurek-03}.  These issues are relevant both
for fundamental reasons, such as to improve our understanding on the
emergence of interference and entanglement useful for
quantum-information purposes~\cite{NielsenChuang} or for enhancing the
efficiency of energy conversion in complex networks~\cite{Lambert-13},
as well as for more applied quantum-thermodynamical purposes, such as
the optimization of energy storage in subportions of the whole
system~\cite{Campaioli-18} or the realization of work extraction
engines at the nanoscale~\cite{BCSW-17}.  The possible presence of
different quantum phases and the development of criticality in the
system may be exploited to characterize the sensitivity to a variety
of behaviors~\cite{Sachdev-book}.

%%%%%%%%%%%%%%%%%%%%%%%%%%%%%%%%%%%%%%%%%%%%%%%%%%%%%%%%%%%%%%%%%%%%
\begin{figure}
  \includegraphics[width=0.95\columnwidth]{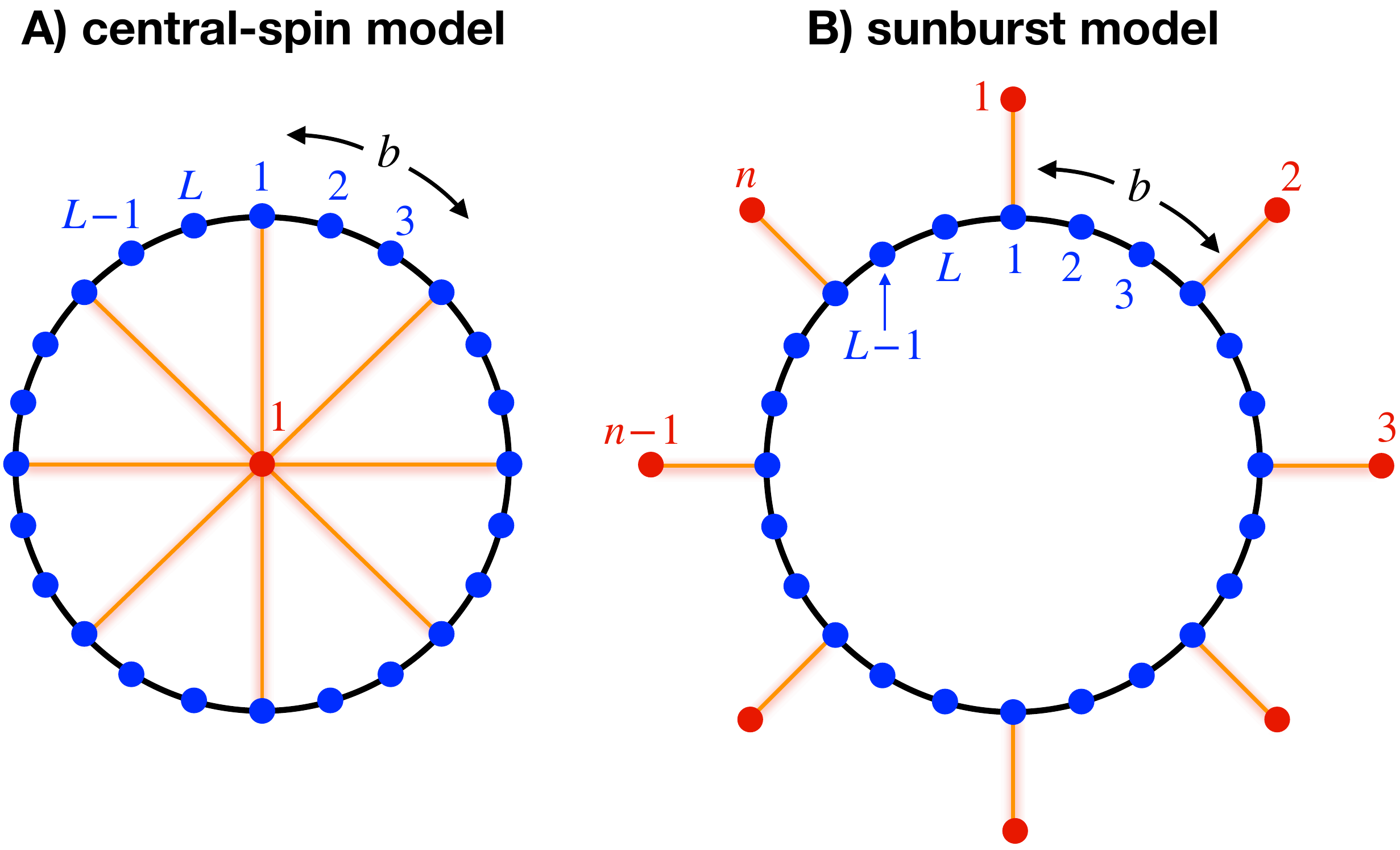}
  \caption{A) Sketch of the ``central-spin model'': a ring made of $L$
    interacting quantum spins (blue dots) is locally coupled to a
    single, central, qubit (red dot) at every $b$ lattice spacings
    ($b=3$ in the figure).  Note that the typical case studied in the
    literature is with $b=1$.  B) Sketch of the ``sunburst model''
    discussed in this paper: a ring of $L$ spins is locally coupled to
    $n$ ancillary qubits at every $b=L/n$ lattice spacings ($L=24$ and
    $n=8$ in the figure).}
  \label{setup}
\end{figure}
%%%%%%%%%%%%%%%%%%%%%%%%%%%%%%%%%%%%%%%%%%%%%%%%%%%%%%%%%%%%%%%%%%%%

From a conceptual point of view, the cleanest scenario is, perhaps,
that of a system composed of several quantum objects described
by a global static Hamiltonian.  A portion
of the system itself can be thus seen as an effective bath for the
remaining part.  In this context, paradigmatic toy models are the
so-called ``central-spin models'' (for a sketch, see
Fig.~\ref{setup}A), where one or few qubits are globally coupled to an
environmental many-body system, see, e.g., Refs.~\cite{Zurek-82,
  CPZ-05, QSLZS-06, RCGMF-07, CFP-07, CP-08, Zurek-09, DQZ-11, NDD-12,
  SND-16, V-18, FCV-19, RV-19, RV-21}.  The main decoherence properties
of a single qubit (red dot) globally coupled to a many-body system
(surrounding blue dots) turn out to crucially depend on the quantum
phase of the many-body system, whether it is within the ordered or
disordered quantum phases, or at the quantum critical point (QCP)
separating them, when it develops long-range correlations~\cite{RV-21}.

Here we come back to issues related to the decoherence arising from
interactions between different parts of an isolated
system and propose an alternative arrangement of its composing quantum
objects, which we call the ``sunburst model'' (a sketch is shown in
Fig.~\ref{setup}B).  Namely, we consider a quantum spin chain (blue
dots) locally coupled to external isolated qubits (surrounding red
dots).  The sunburst model admits a high degree of flexibility in the
choice of the number of external qubits, which can be a finite or an
infinte set at regular spatial intervals, and may be interpreted as a
probing subsystem.  The specific geometry of the model makes it
nonintegrable, apart from very specific cases, therefore numerical
approaches are usually needed to study the emerging physics.

In this paper we study the case in which the inner ring is described
by a quantum Ising chain, being a prototypical quantum many-body
system which presents different quantum phases, with both first-order
and continuous quantum transitions (FOQTs and CQTs, respectively) when
varying the intensity of the two external, transverse and
longitudinal, fields. In this first exploratory study of the sunburst
quantum Ising model, we investigate the equilibrium properties of the
ground state of the global system, focusing on the behavior of
observables associated with the Ising chain or the ancillary qubits
separately. This allows us to address the effects of interactions
between the two parts on the coherence properties of the single part,
and the scaling behaviors associated with the different phases of the
Ising system.  Close to the quantum transitions, we exploit
renormalization-group (RG) and finite-size scaling (FSS)
frameworks~\cite{RV-21}, which constitute the natural theoretical
context to effectively describe the
behavior of systems in proximity of either a CQT~\cite{CPV-14, CPV-15}
or a FOQT~\cite{CNPV-14, PRV-18-fo, RV-18}, as well as in a dynamic
context~\cite{PRV-18-qu, PRV-18-def, PRV-20}, providing the asymptotic
large-size scaling in a variety of situations.  Since here we
  expressly focus on the ground-state properties, we make use of
  the equilibrium FSS formalism~\cite{CPV-14,RV-21}.

We remark that, strictly speaking, decoherence is a dynamic process,
while in this paper we are only discussing the equilibrium
ground-state properties of the sunburst model. Within our FSS
frameworks, this is equivalent to the study of the adiabatic dynamics
of a finite-size system, generally characterized by a nonvanishing
gap, and thus always possible for sufficiently large time scales, even
close to the quantum transitions.

The paper is organized as follows.  In Sec.~\ref{modelobs} we
introduce the various Hamiltonian terms of the sunburst quantum Ising
model, the decoherence functions and all observables that will be
addressed in the subsequent analyses, and discuss some simple limit
cases.  We then study situations where the system parameters are close
to the CQT of the Ising ring, in the large-size limit achieved either
by keeping fixed the number of ancillary qubits coupled to the ring
(Sec.~\ref{fssfn}) or by taking any two consecutive qubits at a fixed
distance (Sec.~\ref{infiniten}).  In Sec.~\ref{foqtbeh} we focus on
the regime close to the FOQT line of the Ising ring, while in
Sec.~\ref{disbeh} we discuss the system behavior within the Ising
disordered phase. Finally, in Sec.~\ref{conclu}, we summarize our
findings and draw conclusions.

\section{The sunburst quantum Ising model}
\label{modelobs}

We start by defining our model, where a quantum Ising ($\mathcal{I}$)
spin-ring is coupled to a set of ancillary ($\mathcal{A}$)
isolated two-level systems, in
a sunburst geometry, as sketched in Fig.~\ref{setup}B.  The quantum
Ising chain is a useful theoretical laboratory where fundamental
issues of quantum many-body systems can be throughly investigated,
exploiting the exact knowledge of several features of its phase
diagram and quantum correlations.  Coupling it to ancillary spins
allows us to define a prototypical situation to address the emerging
critical behavior of composite quantum many-body systems. This is what
we are going to study below.

\subsection{Hamiltonians}
\label{model}

Quantum Ising rings are defined by the Hamiltonian
\begin{equation}
  \hat H_{\cal I} = - \sum_{x=1}^{L} \Big( J \hat \sigma^{(1)}_{x\phantom{1}}
  \hat \sigma^{(1)}_{x+1} + g \hat \sigma^{(3)}_x \Big) \,,
\label{Iring}
\end{equation}
where $L$ is the system size, $\hat \sigma^{(i)}_x$ are the Pauli
matrices on the $x$th site ($i = 1,2,3$ labels the three spatial
directions) and $\hat \sigma^{(i)}_{L+1} = \hat \sigma^{(i)}_1$,
corresponding to periodic boundary conditions (PBC).  In the following
we assume ferromagnetic nearest-neighbor interactions with $J>0$.

Several results for the low-energy properties of model~\eqref{Iring}
have been thoroughly obtained, both in the thermodynamic limit and in
the FSS limit with various boundary conditions (see, e.g.,
Refs.~\cite{Sachdev-book, DMS-book, NO-11, RV-21} and references
therein).  Here we recall that the model undergoes a CQT at $g=g_{\cal I}=J$,
belonging to the two-dimensional Ising universality class, separating
a disordered phase ($g>g_{\cal I}$) from an ordered one ($g<g_{\cal I}$).
Approaching the CQT, the system develops long-distance correlations,
with a length scale $\xi$ diverging as $\xi \sim |g-g_{\cal I}|^{-\nu}$,
where $\nu=y_g^{-1}=1$ and $y_g$ is the RG dimension associated with
the difference $g-g_{\cal I}$. The ground-state energy gap $\Delta_{\cal I}$
(i.e., the energy difference between the two lowest Hamiltonian
eigenstates of $\hat H_{\cal I}$) gets suppressed as
$\Delta_{\cal I} \sim \xi^{-z}$,
where $z$ is the dynamic critical exponent ($z=1$). In particular, at
the critical point~\cite{BG-85},
\begin{equation}
  \Delta_{\cal I}(g_{\cal I},L) = {\pi \over 2L} + O(L^{-2})\,.
  \label{fssdelta}
\end{equation}
Another independent critical exponent arises from the RG dimension of
the symmetry-breaking homogeneous longitudinal field $h$, coupled to
an additional Hamiltonian term of the form
$\sum_{x=1}^L \hat \sigma_x^{(1)}$, which is $y_h=(2+d+z-\eta)/2=2-\eta/2$,
where $d$ stands for the system dimensionality (here $d=1$) and $\eta=1/4$,
thus $y_h=15/8$ (see, e.g., Refs.~\cite{Sachdev-book, RV-21}).
Along the $|g/J|<1$ line, the longitudinal field $h$ drives FOQTs,
associated with the avoided level crossing of quantum states with
opposite magnetization in finite-size systems, which leads to a
discontinuity of the longitudinal magnetization $M=L^{-1} \big\langle
\sum_x \hat \sigma_x^{(1)} \big \rangle$ in the infinite volume limit.
The energy difference of the lowest states along the
FOQT line of the Ising ring is exponentially suppressed with
increasing $L$ as~\cite{Pfeuty-70, CJ-87}
\begin{equation}
  \Delta_{\cal I}(g<g_{\cal I},L) \approx 2 \,\sqrt{g_{\cal I}^2-g^2\over \pi L} \; (g/J)^L \,.
    \label{deltapbc}
\end{equation}

In our sunburst Ising model, we assume to have $n$ ancillary isolated
spin-$1/2$ systems (qubits), each of them coupled with one spin of the
Ising ring, located at an equal distance $b$ each other, so that
$b=L/n$ (see Fig.~\ref{setup}B).  We assume they are all described by
the same Hamiltonian
\begin{equation}
  \hat H_{\cal A} = - {\delta \over 2} \sum_{j=1}^n \hat \Sigma_j^{(3)} \,,
  \label{HA}
\end{equation}
where $\hat \Sigma_j^{(i)}$ are the Pauli matrices of the $j$th
ancillary qubit.  The parameter $\delta$ quantifies the energy
difference between the lowest eigenstates of the
Hamiltonian~\eqref{HA}, thus providing the energy scale associated
with the ancillary system.

The global closed system is formed by two subsystems: the Ising ring
with $L$ spins and the $n$ ancillary two-level systems. Their
interaction is assumed to be homogeneous for all $n$ qubits and
described by the Hamiltonian coupling
\begin{equation}
  \hat H_{{\cal I}{\cal A}} = - \kappa \sum_{j=1}^n \hat \Sigma_j^{(1)} \hat
  \sigma_{x=jb}^{(1)}\,,
  \label{HIA}
\end{equation}
with a strength controlled by the parameter $\kappa$.

The full sunburst-Ising-model Hamiltonian
\begin{equation}
  \hat H = \hat H_{\cal I} + \hat H_{\cal A} + \hat H_{{\cal I}{\cal A}}
  \label{globha}
\end{equation}
maintains some of the invariance under translation, i.e., the
translation of $b$ lattice spacings. Moreover, the global system has a
${\mathbb Z}_2$ symmetry, which also involves the ancillary spin
operators:
\begin{eqnarray}
  & \hat \sigma_x^{(1/2)} \to - \hat \sigma_x^{(1/2)}\,,\qquad
  &\hat \Sigma_j^{(1/2)} \to - \hat \Sigma_j^{(1/2)}\,,\label{z2symm}\\
    & \hat \sigma_x^{(3)} \to \hat \sigma_x^{(3)}\,,\qquad
  &\hat \Sigma_j^{(3)} \to \hat \Sigma_j^{(3)}\,.\nonumber
\end{eqnarray}
Of course, the ${\mathbb Z}_2$ symmetry associated with the bare Ising
chain, i.e., 
\begin{eqnarray}
  \hat \sigma_x^{(1/2)} \to - \hat \sigma_x^{(1/2)}\,,\qquad
\hat \sigma_x^{(3)} \to \hat \sigma_x^{(3)}\,,
\label{z2symmisi}
\end{eqnarray}
does not hold anymore. Moreover, one can easily show that the phase
diagram, and generic global observables, are symmetric with respect to
changes of the sign of $\kappa$ and/or $\delta$.  Without loss of
generality, in the following we restrict ourselves to $\kappa\ge 0$
and $\delta\ge 0$; we also set $J = \hbar = 1$, thus providing the
corresponding unities for all the other quantities.

We finally note that, for $n=L$, the above sunburst model may be
related to the so-called Heisenberg Ising-Kondo necklace model, for
some values of its parameters (see, e.g., Ref.~\cite{RC-18}).

\subsection{Decoherence functions and observables}
\label{obs}

We address the behavior of both subsystems composing the sunburst
model (the Ising ring and the ancillary isolated qubits), within the
global ground state $|\Psi_0\rangle$. In particular, we aim at
studying how their main features, such as the quantum phases and the
critical behavior, change when varying the Hamiltonian parameters.  To
this purpose, we first consider the reduced density matrices
associated with the Ising ring ($\rho_{\cal I}$) and the ancillary
subsystem ($\rho_{\cal A}$) corresponding to the ground state of the
global system,
\begin{equation}
  \rho_{\cal I} = {\rm Tr}_{\cal A} \, \big[ |\Psi_0\rangle
    \langle\Psi_0| \big] \,, \qquad \rho_{\cal A} = {\rm Tr}_{\cal I}
  \, \big[ |\Psi_0\rangle \langle\Psi_0| \big] \,,
  \label{rhoIA}
\end{equation}
where ${\rm Tr}_{\cal A} [\,\cdot\,]$ and ${\rm Tr_{\cal I}} [\,\cdot\,]$ denote
the partial traces over the respective counterparts.

Their coherence properties can be quantified through the purities
$P_{\cal I}$ and $P_{\cal A}$, defined as
\begin{equation}
  P_{\cal I} = {\rm Tr} \big[ \rho_{\cal I}^2 \big] \,,\qquad
  P_{\cal A} = {\rm Tr} \big[ \rho_{\cal A}^2 \big] \,,
  \label{purity}
\end{equation}
or, equivalently, by the decoherence factor
\begin{equation}
  Q_{\cal I} = 1 - P_{\cal I}\,, \qquad Q_{\cal A} = 1 - P_{\cal A}\,.
  \label{Qdef}
\end{equation}
One may also define the corresponding R{\'e}nyi entanglement entropies
$S_{\cal I}$ and $S_{\cal A}$, as
\begin{equation}
  S_{\cal I} = - {\rm ln} \, P_{\cal I} \,,\qquad
  S_{\cal A} = - {\rm ln} \, P_{\cal A} \,.
  \label{enten}
\end{equation}
Exploiting the Schmidt decomposition for bipartitions
of pure states~\cite{NielsenChuang}, one can easily prove that
\begin{equation}
  P_{\cal I} = P_{\cal A} \equiv P\,, \qquad Q_{\cal I} = Q_{\cal A} \equiv Q\,,
  \qquad S_{\cal I} = S_{\cal A} \equiv S\,.
  \label{pia}
\end{equation}
The decoherence factor is limited between $Q=0$ (corresponding to
$P=1$ and $S=0$, for a pure reduced state) and $Q\to 1$ (corresponding
to $P \to 0$, for a completely incoherent many-body state).  Now,
since $Q$ must be an even function of $\kappa$ and assuming
analyticity at $\kappa=0$, we may expand it at small $\kappa$ as
\begin{equation}
  Q \approx {1\over 2} \kappa^2 \chi_Q + O(\kappa^4)\,, \qquad
  \chi_Q\equiv {\partial^2 Q\over \partial \kappa^2}\bigg|_{\kappa=0}\,,
\label{chiqdef}
\end{equation}
where $\chi_Q$ keeps the role of ``decoherence susceptibility''.

To characterize the effects of interactions between the two
subsystems, we also consider other indicators and observables.  As a
global quantity, we study the gap $\Delta$, namely the energy
difference between the first excited state and the ground state of the
full system.
Then we focus on observables related to the Ising ring only, such as
correlations of the spin operators $\hat \sigma_x^{(1)}$.  Due to the
${\mathbb Z}_2$ symmetry~\eqref{z2symm}, ${\rm Tr} \big[\rho_{\cal I}
  \hat{\sigma}_{x}^{(1)} \big]=0$.  The two-point correlation function
can be written as
\begin{equation}
  G(x,y) = {\rm Tr}
  \big[ \rho_{\cal I} \hat\sigma_{x}^{(1)} \hat\sigma_{y}^{(1)} \big]
  \,.\label{gxy}
\end{equation}
Defining its Fourier-like transform along the ring as
\begin{equation}
  \widetilde{G}(p) = \frac{1}{L} \sum_{x,y} e^{-ip(x-y)} G(x,y)\,,
  \label{gip}
\end{equation}
we consider its zero-momentum component $\chi = \widetilde{G}(0)$ and
second-moment correlation length
\begin{equation}
  \xi^2 = {1\over 4 \sin^2(\pi/L)}
     {\widetilde{G}(0) - \widetilde{G}(2\pi/L)
       \over \widetilde{G}(2\pi/L}\,,
\label{xidef}
\end{equation}
where $2\pi/L$ is the minimum finite momentum along the ring,
representing a natural choice of an $O(1/L)$ nonzero momentum for the
Ising subsystem.  Finally we also define a Binder-like parameter
\begin{equation}
  U = {{\rm Tr}[\rho_{\cal I} \hat \mu_2^2]\over ({\rm Tr}[\rho_{\cal I}
      \hat\mu_2])^2}\,, \quad \mbox{with} \quad \hat \mu_2 = {1\over L^2}
  \sum_{xy}\hat\sigma^{(1)}_x\hat\sigma_y^{(1)}\,.
\label{Udef}
\end{equation}

\subsection{Extreme cases and notable limits}
\label{limits}

To understand the role of the Hamiltonian parameters $\delta$ and
$\kappa$, respectively describing the ancillary spin system (${\cal
  A}$) and the interaction with the Ising ring (${\cal I}$)
[cf.~Eqs.~\eqref{HA}-\eqref{HIA}], it is first useful to discuss some
particularly simple cases:
\begin{enumerate}
  \item When $\kappa=0$, the two subsystems ${\cal A}$ and ${\cal I}$
    are trivially decoupled, therefore $|\Psi_0\rangle = |
    A_0\rangle_{\cal A} \otimes | \psi_0\rangle_{\cal I}$ where
    $|A_0\rangle_{\cal A}$ and $| \psi_0\rangle_{\cal I}$ denote the ground states
    of the isolated qubits and of the Ising ring.
    
  \item When $\delta=0$, the lowest levels of the global system ${\cal
    A}+{\cal I}$ turn out to be twofold degenerate.

  \item When $\delta\to\infty$, the state of each of the isolated
    qubits gets fixed to the eigenstate of $\hat \Sigma^{(3)}$ with
    eigenvalue one, which can also be written as a superposition of
    the eigenstates $|\pm \rangle$ of $\hat \Sigma^{(1)}$.  The global
    ground state is thus again given by the product of states of the
    two subsystems, i.e., $|\Psi_0\rangle = | A_0\rangle_{\cal A}
    \otimes | \psi_0\rangle_{\cal I}$, where $| A_0\rangle_{\cal A} =
    |\phi_0\rangle_1 \otimes \cdots \otimes |\phi_0\rangle_n$, with
    $|\phi_0\rangle_j = (|+\rangle_j + |-\rangle_j)/\sqrt{2}$ for the
    $j$th ancillary qubit, and $|\psi_0\rangle_{\cal I}$ is the ground
    state of the closed Ising ring (note that the interaction term
    associated with $\kappa$ vanishes on the ancillary state
    $|A_0\rangle_{\cal A}$).

  \item The $\kappa\to\infty$ limit is best discussed without
    fixing $J=1$, i.e., leaving it free.  When $\kappa\to\infty$, the
    spins of the Ising rings coupled with the ancillary system are
    forced to stay in a state described by the reduced density matrix
    \begin{subequations}
    \begin{equation}
      \qquad \rho_{\cal A} = \frac{1}{2} \bigg(
      \bigotimes_{j=1}^n |+\rangle_j \langle +|
      \, + \, \bigotimes_{j=1}^n|-\rangle_j \langle -| \bigg) 
    \end{equation}
    in the limit $J\to \infty$ in the Hamiltonian~\eqref{Iring}, and
    \begin{equation}
      \qquad \rho_{\cal A} = \bigotimes_{j=1}^n \frac12 \bigg(
      |+\rangle_j \langle +| \, + \, |-\rangle_j \langle -| \bigg)
    \end{equation}
    \end{subequations}
    in the limit $J\to 0$.  As a consequence the corresponding
    decoherence measure of the Ising ring ranges from $Q = 1/2$ (for
    $J\to \infty$) to $Q=1-1/2^n$ (for $J\to 0$).
\end{enumerate}

\begin{figure}[!t]
    \includegraphics[width=0.95\columnwidth]{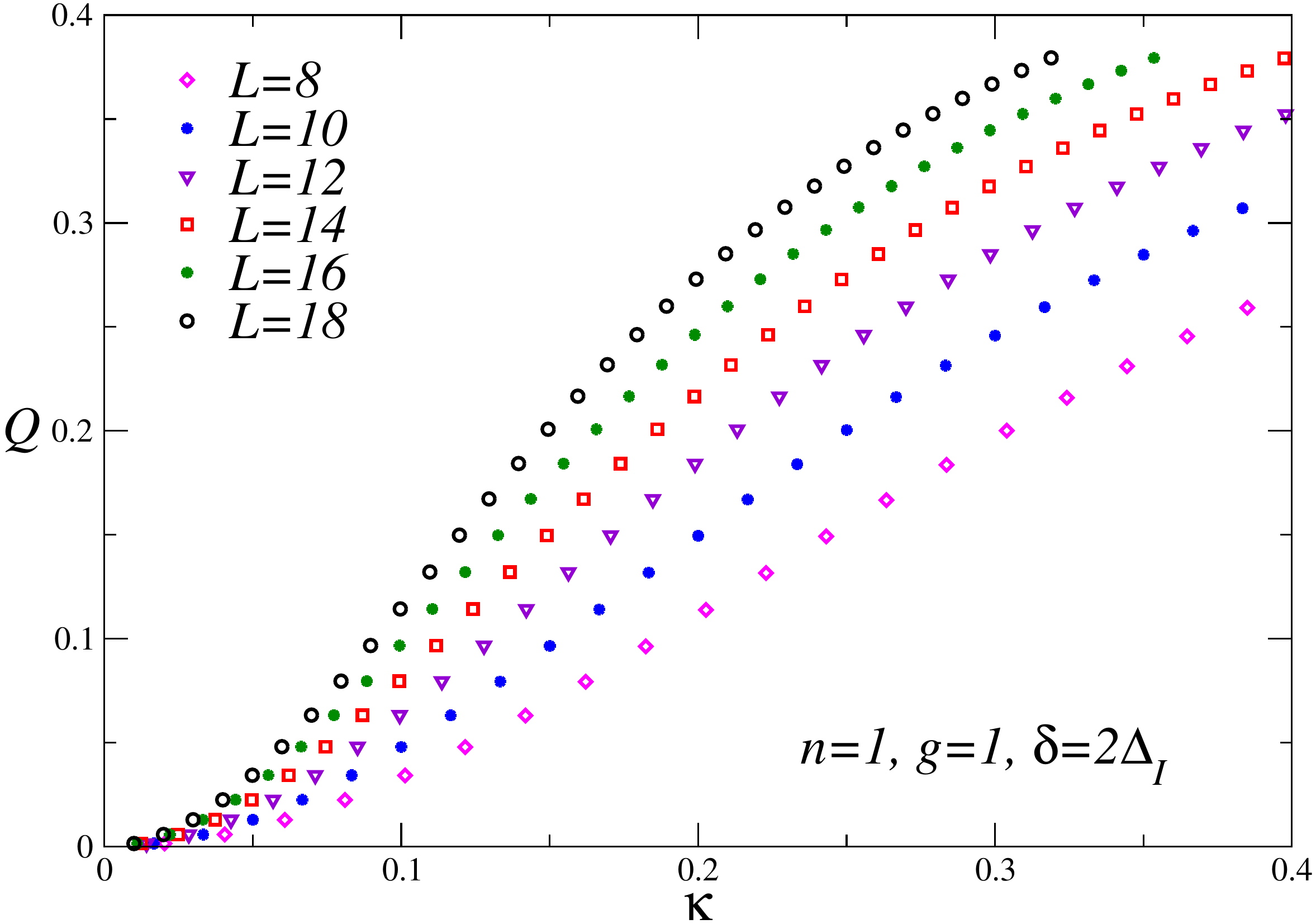}
    \includegraphics[width=0.95\columnwidth]{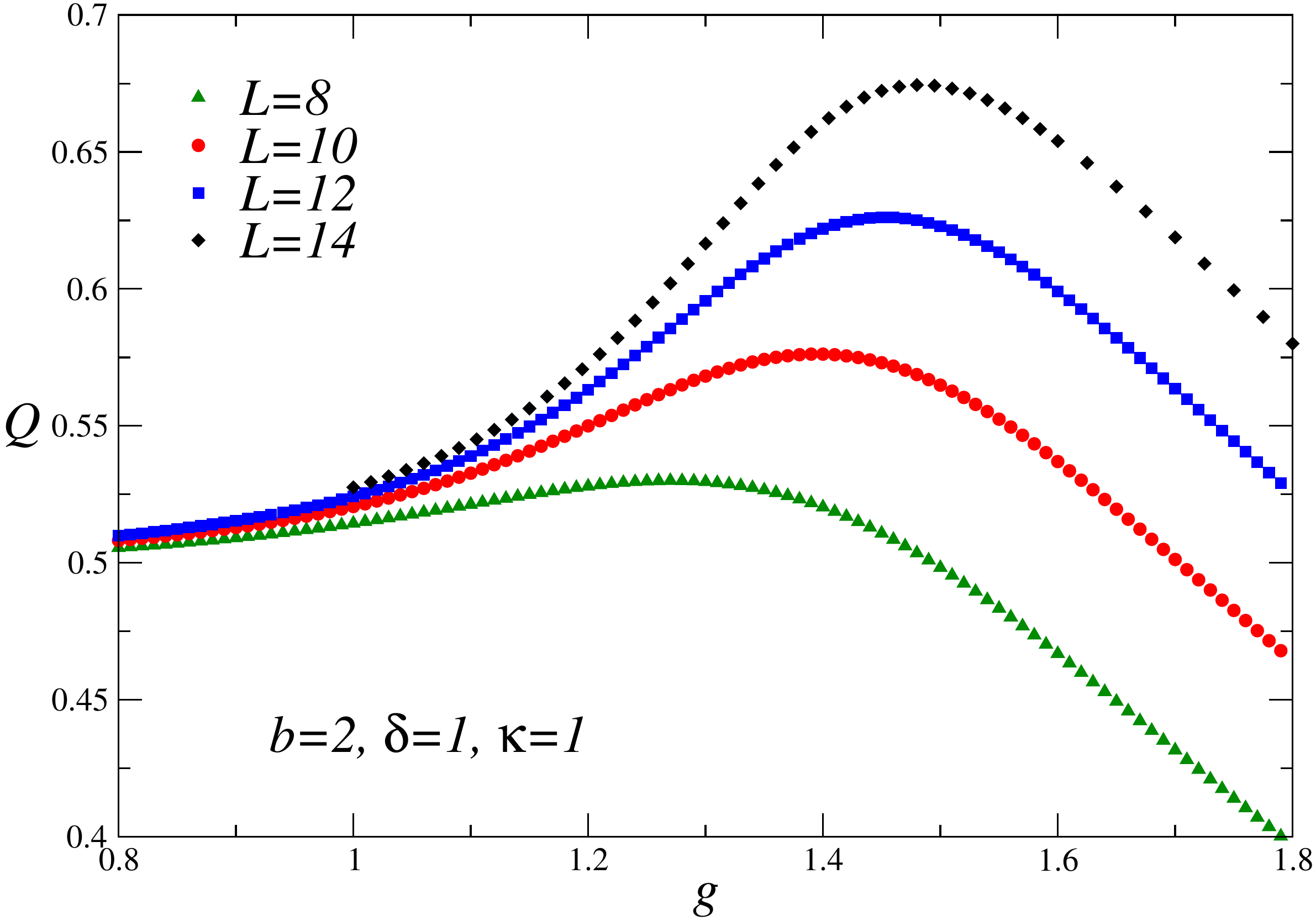}
    \caption{Behavior of the decoherence factor $Q$ as a function of
      the various parameters of the sunburst Ising model,
      for different chain lengths $L$.  Top panel: $Q$ vs.~$\kappa$,
      for $g=g_{\cal I}=1$, $\delta = 2 \, \Delta_{\cal I}(g_{\cal I}, L)$,
      and fixing the number of ancillary spins ($n=1$).
      Bottom panel: $Q$ vs.~$g$, for $\delta = \kappa = 1$, and fixing
      the distance between two consecutive ancillary spins ($b=2$).}
    \label{figQexample}
\end{figure}

The above considerations hint at the fact that the decoherence effects
of the ancillary system may significantly depend on the sunburst-model
parameters, as explicitly shown in Fig.~\ref{figQexample} (data have
been obtained by means of numerical exact diagonalization techniques).
The top panel shows how the decoherence factor $Q$ sets in when the
coupling $\kappa$ between the critical Ising ring and the ancillary
qubits increases, keeping the ratio $\delta/\Delta_{\cal I}$ of their energy
scales fixed; here the thermodynamic limit is taken by fixing the
number $n$ of qubits in the sunburst geometry [case (I) below]. We
observe significant differences when increasing the ring size.  The
bottom panel of Fig.~\ref{figQexample} displays the behavior of the
decoherence factor $Q$, for finite values of $\delta$ and $\kappa$, as
a function of the transverse field strength $g$ (which drives the
critical behavior on the Ising ring itself). Here we approach the
large-size limit by keeping the distance $b$ between two consecutive
ancillary spins fixed and then increasing the chain length $L$ [case
  (II) below].  We observe the emergence of drastic changes in the
decoherence properties of the lattice with $g$, signaling that
different phases are related to qualitative variations of the
decoherence factor $Q$.

%%%%%%%%%%%%%%%%%%%%%%%%%%%%%%%%%%%%%%%%%%%%%%%%%%%%%%%%%%%%%%%%%%%%
%%%%%%%%%%%%%%%%%%%%%%%%%%%%%%%%%%%%%%%%%%%%%%%%%%%%%%%%%%%%%%%%%%%%

In the remaining part of this paper we want to study the behavior of
the sunburst Ising model~\eqref{globha} for generic values of $\delta$
and $\kappa$, in the limit of large size, by exploiting RG and FSS
frameworks.  In this respect, we consider two different situations
where the large-$L$ limit can be achieved:
\begin{enumerate}
\item[(I)] We keep the number $n$ of the ancillary spins finite and
  fixed, while $L$ is increased (same situation as in the top panel of
  Fig.~\ref{figQexample}).  In this case, the bulk properties of the
  Ising ring are expected to remain unchanged.  Thus we expect that,
  for any value of Hamiltonian parameters $\kappa$ and $\delta$, the
  system develops an Ising-like CQT at $g=g_{\cal I}$ characterized by
  Ising critical modes along the Ising ring, separating disordered
  and ordered phases.

\item[(II)] We keep the distance $b=L/n$ between two consecutive
  ancillary spins fixed, such that their number increases linearly
  with $L$ (same as in the bottom panel of Fig.~\ref{figQexample}).
  In this case the ancillas are expected to give rise to drastic
  changes, with transition lines related to the breaking of the global
  ${\mathbb Z}_2$ symmetry. In particular, such transitions are
  expected to move the critical point $g=g_{\cal I}$ of the closed Ising
  ring, generally to $g_c>g_{\cal I}$, as we shall see later.
\end{enumerate}

\section{Sunburst Ising model at the CQT, with a finite number of qubits}
\label{fssfn}

We start to discuss the effects of a finite number $n$ of isolated
qubits coupled with the Ising ring, as described by the
Hamiltonian~\eqref{globha}, thus the distance between them increases
as $b=L/n$ in the large-size limit [case (I) above].  The ancillary
spins play the role of particular defects inserted within the bulk of
the Ising system, controlled by the parameter $\kappa$,
cf. Eq.~\eqref{HIA}.

As already mentioned, for an ancillary system with a finite number $n$
of isolated spins, the bulk properties of the Ising ring do not change
qualitatively, still showing disordered and ordered phases separated
by the CQT at $g=g_{\cal I}=1$, independently of the parameters $\kappa$
and $\delta$. However, some notable change emerges in the FSS behavior
close the CQT, and of course also in the coherence properties of the
Ising ring.
In particular, we distinguish between two different FSS regimes:\\
(i) a scaling limit where the interaction with the ancillary spin-system
Hamiltonian acts as a perturbation, thus both $\kappa$ and $\delta$ are small,
and scale appropriately with increasing the size of the Ising ring
(cf.~Sec.~\ref{smallkl});\\
(ii) a FSS limit at fixed values of $\delta>0$ (cf.~Sec.~\ref{finitekl}).

\subsection{FSS around the closed Ising-ring criticality}
\label{smallkl}

The first question we address is how perturbations arising from the
interaction with the ancillary system affect the critical behavior of
the closed Ising ring at $g\approx g_{\cal I}$. When the Hamiltonian
$\hat H_{{\cal I}{\cal A}}$, cf.~Eq.~\eqref{HIA}, acts as a perturbation, we
may derive FSS laws using RG arguments based on the scaling behavior
of the parameters $\delta$ and $\kappa$.  To address this problem, we
should first introduce scaling variables associated with them.

The standard FSS of the closed Ising ring at the CQT is controlled
by the scaling variable
\begin{equation}
  W = (g-g_{\cal I})L^{y_g} \,, \qquad y_g=1\,.
  \label{wdef}
\end{equation}
The parameter $\delta$ represents the energy scale associated with the
isolated qubits. Therefore, its scaling variable should be provided by
the ratio between it and the gap $\Delta_{\cal I}(g_{\cal I},L)$
of the Ising ring at criticality, i.e.,
\begin{equation}
  A = \delta/\Delta_{\cal I} \sim \delta \, L^z \,, \qquad z=1\,.
  \label{adef}
\end{equation}
To identify the scaling variable associated with the parameter
$\kappa$, we note that the corresponding perturbations may be
interpreted as symmetry-breaking defects for the Ising ring [under the
$\mathbb{Z}_2$ transformation~\eqref{z2symmisi} restricted to the
Ising-ring subsystem only], controlled by an external operator
$\hat\Sigma_j^{(1)}$, i.e.,
\begin{equation}
  \hat D_x = -\kappa \, \hat\Sigma_j^{(1)} \hat\sigma_{x=jb}^{(1)}\,, 
    \label{pkappa}
\end{equation}
where the index $x$ denotes the position of the defect along the
ring.  Since $\hat\Sigma_j^{(1)}$ is related to the noncritical
ancillary system, we expect it to be not relevant for the
determination of the RG dimension of the parameter
$\kappa$. Therefore, it should be the same as that of the simpler
symmetry-breaking defect
\begin{equation}
  \hat D_x = -\kappa \, \hat\sigma_x^{(1)}
  \label{dxdef}
  \end{equation}
  for the critical Ising ring, which is equal to
  $y_\kappa=7/8$~\cite{FRV-22}.  The scaling variable associated with
  $\kappa$ should be thus given by
\begin{equation}
  K = \kappa L^{y_\kappa} \,, \qquad y_\kappa=7/8\,.
\label{kdef}
\end{equation}
In summary, the FSS limit of all the various observables introduced in
Sec.~\ref{obs} can be defined as the large-size limit keeping $W$,
$A$, and $K$ fixed, therefore for $\delta\sim L^{-z}$ and $\kappa \sim
L^{-y_\kappa}$.

%%%%%%%%%%%%%%%%%%%%%%%%%%%%%%%%%%%%%%%%%%%%%%%%%%%%%%%%%%%%%%%%%%%%
\begin{figure}[!t]
  \includegraphics[width=0.95\columnwidth]{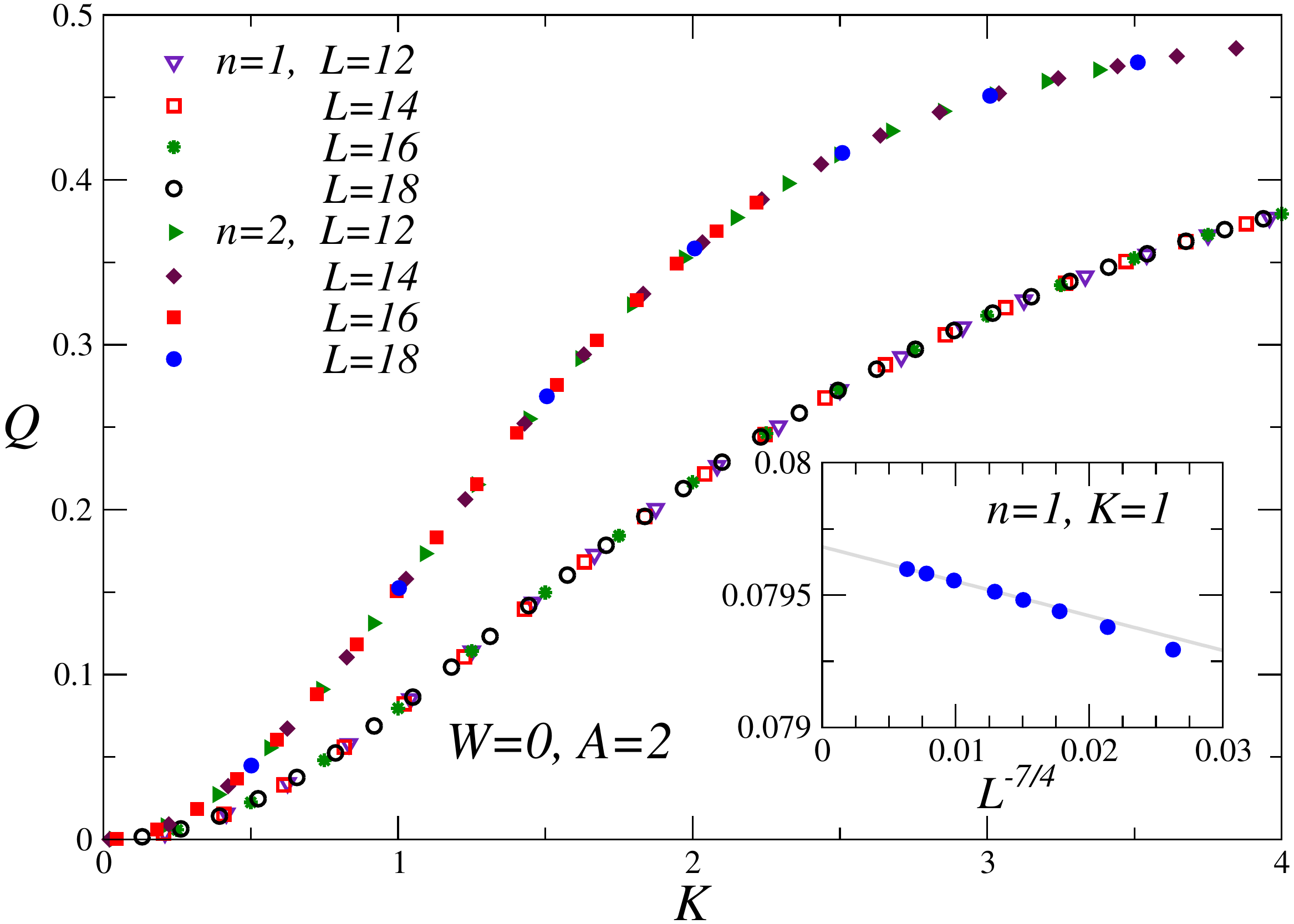}
  \includegraphics[width=0.95\columnwidth]{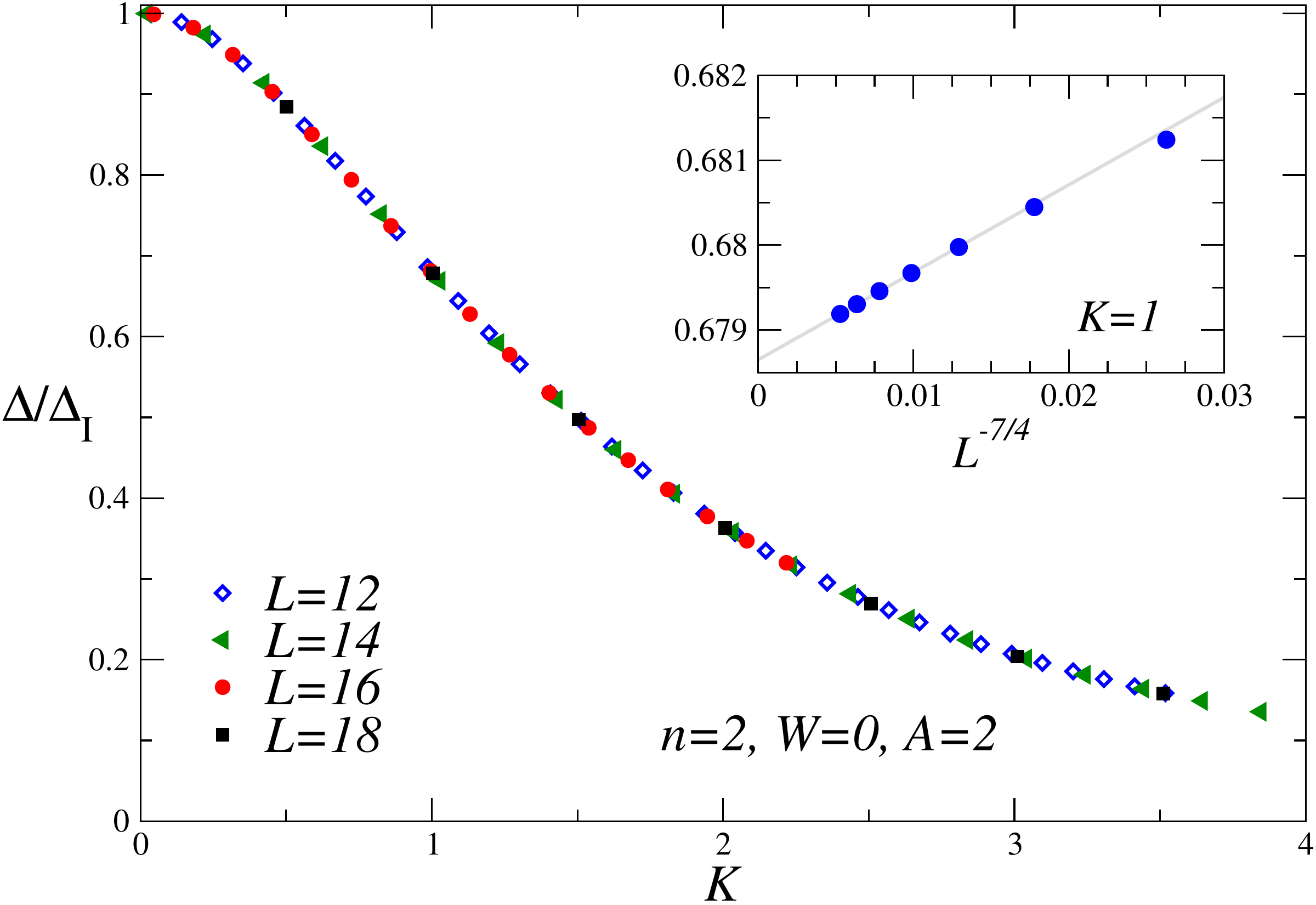}
  \includegraphics[width=0.95\columnwidth]{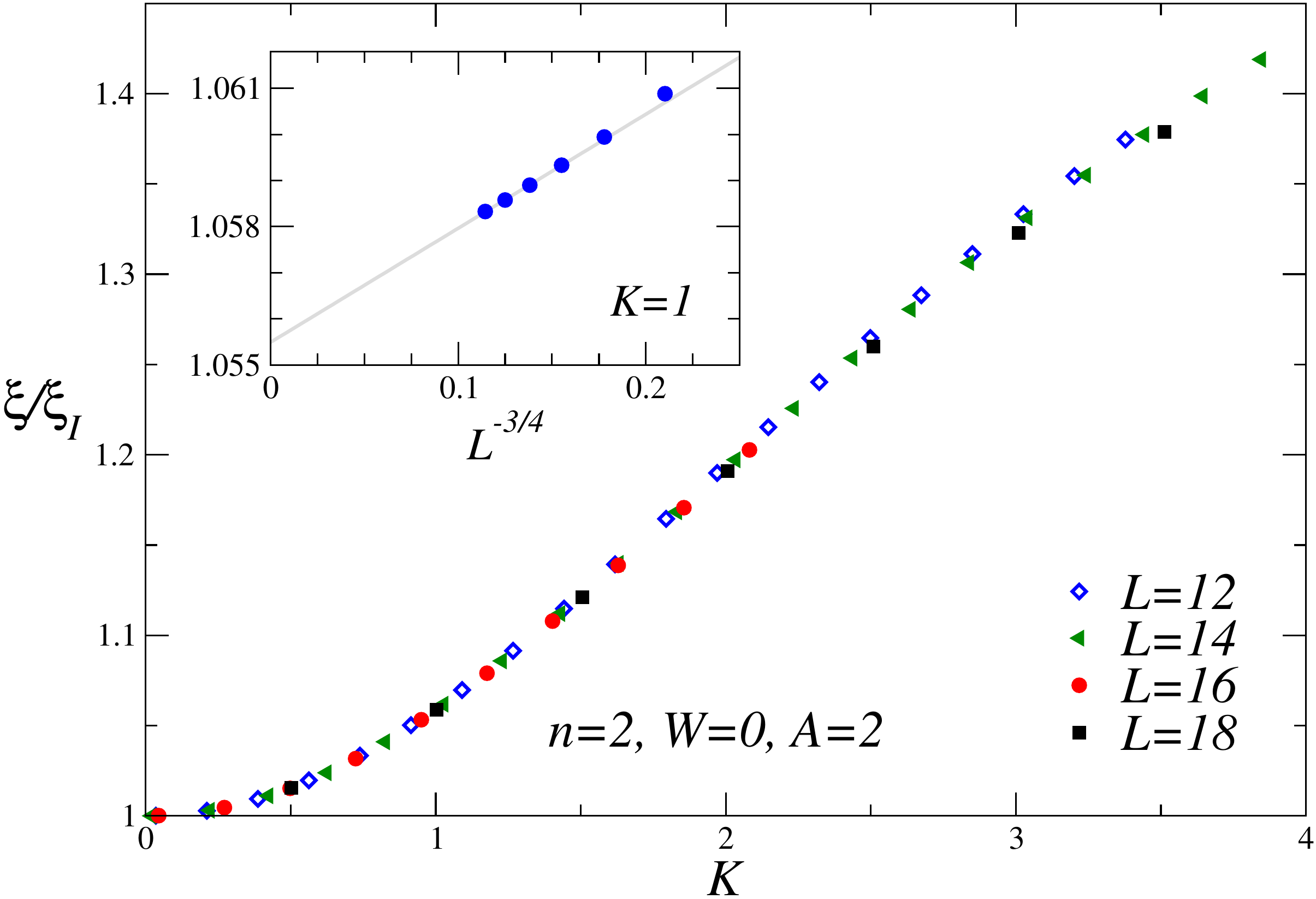}
  \caption{Scaling of the decoherence factor $Q$ (top panel), the
    ratio of the gaps $\Delta/\Delta_{\cal I}$ defined in Eq.~\eqref{gapsca2}
    (central panel), and the ratio of the correlation lengths
    $\xi/\xi_{\cal I}$ defined in Eq.~\eqref{ratioconv} (bottom panel) in
    terms of the variable $K=\kappa L^{7/8}$.  The various data sets
    are for different sizes of the Ising ring and for
    a fixed number $n$ of isolated qubits. We also fix $W=0$ and
    $A=2$.  They confirm the predicted scaling behaviors, showing
    convergence to asymptotic large-$L$ FSS curves.  The
    insets display corrections to the scaling of the
    corresponding quantities at fixed $K=1$, which are compatible with
    a decay law $L^{-7/4}$ for $Q$ and $\Delta/\Delta_{\cal I}$, and with
    $L^{-3/4}$ for $\xi/\xi_{\cal I}$ (straight lines interpolate the points
    for the largest available sizes at our disposal and are drawn to
    guide to the eye).}
  \label{figQscaling}
\end{figure}
%%%%%%%%%%%%%%%%%%%%%%%%%%%%%%%%%%%%%%%%%%%%%%%%%%%%%%%%%%%%%%%%%%%%

To monitor the decoherence properties of the Ising ring and the
ancillary subsystem, we consider their decoherence factor $Q$ defined
in Eqs.~\eqref{Qdef}. A natural ansatz for its scaling behavior in the
small $\kappa$ and $\delta$ regime at the CQT of the Ising ring is
provided by the scaling equation
\begin{equation}
  Q(n,g,\kappa,\delta,L) \approx {\cal Q}(n,W,K,A)\,.
  \label{puritysca}
\end{equation}
Note that
\begin{equation}
  {\cal Q}(n,W,-K,A)={\cal Q}(n,W,K,A)\,,
  \label{parityprop}
\end{equation}
and ${\cal Q}(n,W,0,A)=0$.  The scaling in Eq.~\eqref{puritysca} implies
that the corresponding susceptibility $\chi_Q$, cf.~Eq.~\eqref{chiqdef},
scales as
\begin{equation}
  \chi_Q(n,g,\delta,L) \propto \left( {\partial K\over \partial
  \kappa}\right)^2 {\cal C}(n,W,A) = L^{2y_\kappa} {\cal C}(n,W,A)\,.
\label{chiqsca}
\end{equation}
Therefore the decoherence susceptibility is expected to develop a
nonanalytic power-law divergent behavior in the small-$\delta$ regime.

\begin{figure}[!t]
    \includegraphics[width=0.95\columnwidth]{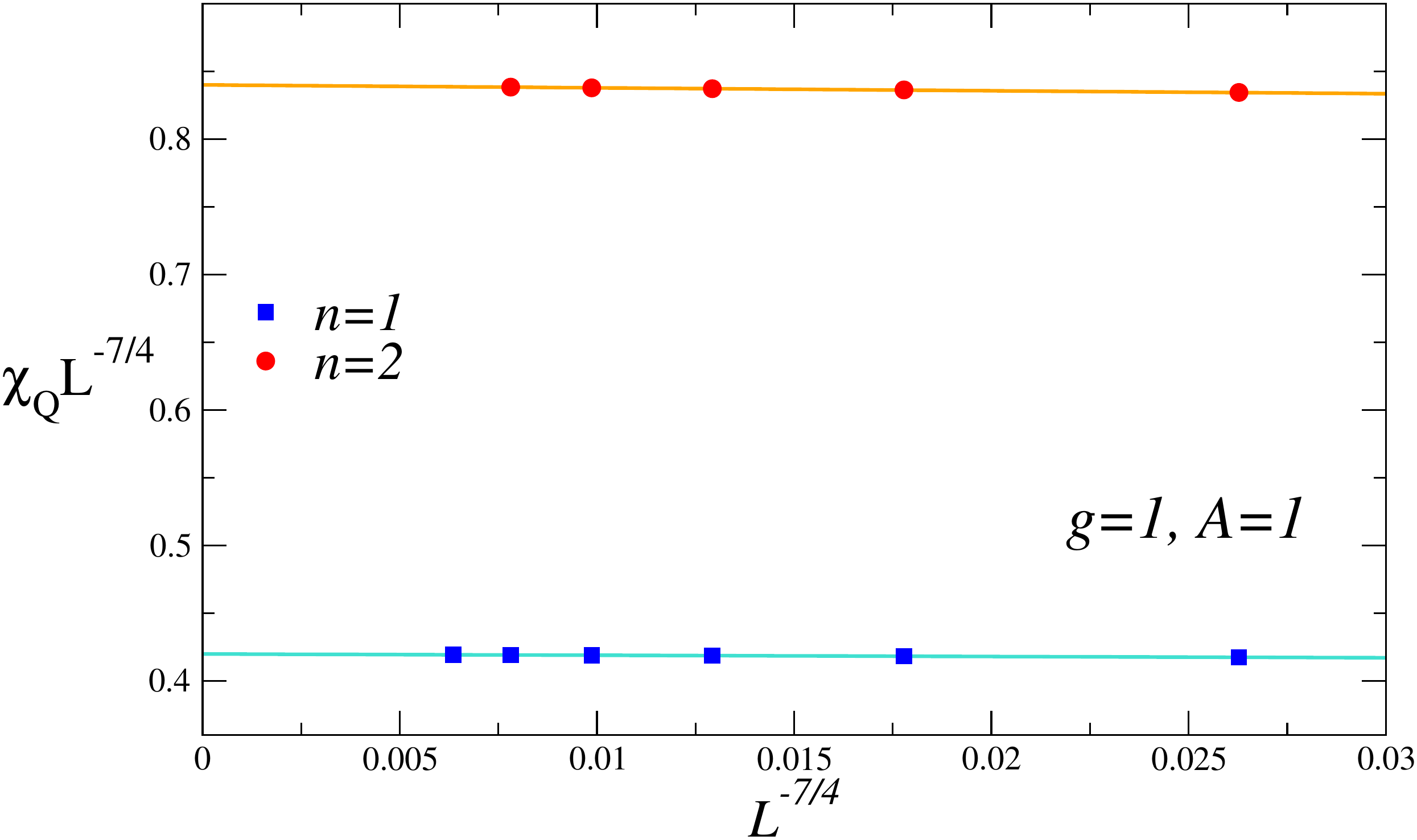}
    \caption{Rescaled decoherence susceptibility $\chi_Q\, L^{-7/4}$
      versus $L^{-7/4}$ (that is, the way in which scaling corrections
      are expected to be suppressed), for two values of $n$ and for
      fixed $W=0$, $A=1$.  Systematic errors stemming from the
      discretization of the second derivative are smaller than the
      marker sizes~\cite{corrQ1n2n}.  Straight lines are drawn to
      guide the eye. The data confirm the asymptotic power-law
      divergence $\chi_Q\sim L^{2y_\kappa}= L^{7/4}$ predicted by the
      scaling Eq.~\eqref{chiqsca}.}
    \label{figsuscQ1n2n}
\end{figure}

We emphasize that the observation of the above FSS laws requires that
all Hamiltonian parameters $g$, $\delta$, and $\kappa$ must be
properly rescaled.  In fact, it is easy to verify that the decoherence
factor with respect to the parameter $\kappa$ does not present
signatures of data collapse under increasing the Ising lattice size
$L$ [see the top panel of Fig.~\ref{figQexample}, where we used the
  scaling variables $W=0$, $A=2$, but we did not scale $\kappa$
  according to the scaling variable $K$ defined in Eq.~\eqref{kdef}].
However, the scaling hypothesis put forward in Eq.~\eqref{puritysca}
is in complete agreement with the data reported in the top panel of
Fig.~\ref{figQscaling}, where $Q$ is plotted in terms of the correct
scaling variable $K$.  It is also possible to verify that the
decoherence susceptibility supports a power-law divergence with $L$,
as outlined in Eq.~\eqref{chiqsca} [see data reported in
  Fig.~\ref{figsuscQ1n2n}, for $n=1$ and $2$].

%%%%%%%%%%%%%%%%%%%%%%%%%%%%%%%%%%%%%%%%%%%%%%%%%%%%%%%%%%%%%%%%%%%%
%%%%%%%%%%%%%%%%%%%%%%%%%%%%%%%%%%%%%%%%%%%%%%%%%%%%%%%%%%%%%%%%%%%%

The scaling behavior of the gap of the global system is expected to be
\begin{equation}
  \Delta(n,g,\kappa,\delta,L) \approx L^{-z} {\cal D}(n,W,K,A)\,.
  \label{gapsca}
\end{equation}
Therefore, the ratio of the global gap with the gap of the closed
Ising ring at the QCP is expected to behave as
\begin{equation}
  {\Delta(n,g=g_{\cal I},\kappa,\delta,L) \over \Delta_{\cal I}(g_{\cal I},L)}
  \approx {\cal D}_c(n,K,A)\,,
  \label{gapsca2}
\end{equation}
as shown in the central panel of Fig.~\ref{figQscaling}.

The two-point function~\eqref{gxy} should behave analogously, apart
from an overall power law, i.e.,
\begin{equation}
  G(x,y,n,g,\kappa,\delta,L) \approx L^{-2y_\varphi}{\cal G}(n,X,Y,W,K,A)\,,
  \label{gxisca}
\end{equation}
where $y_\varphi$ is the RG dimension of the order-parameter operator
at Ising transitions, i.e.
\begin{equation}
  y_\varphi= (d+z-2+\eta)/2 = 1/8\,,
  \label{tvarphi}
\end{equation}
and $X=x/L$, $Y=y/L$ (note that, since the couplings with the ancillary
spins break the translation invariance, we must keep the dependence on
both spatial variables).  In particular, the corresponding
susceptibility $\chi = \widetilde{G}(0)$ should scale as
\begin{equation}
  \chi(n,g,\kappa,\delta,L) \approx L^{1-\eta}{\cal
    B}(n,W,K,A)\,.
  \label{chisca}
\end{equation}
The RG-invariant quantities (not subject by overall power laws), such
as the ratio $R_\xi=\xi/L$, and the Binder parameter $U$,
cf.~Eqs.~\eqref{xidef} and~\eqref{Udef}, are expected to behave as (we
denote them with $R$)
\begin{equation}
  R(n,g,\kappa,\delta,L)  \approx {\cal R}(n,W,K,A)\,.
  \label{rsca}
\end{equation}
Results for the scaling of the RG-invariant ratio
\begin{equation}
  {\xi(n, g=g_{\cal I},
    \kappa, \delta, L)\over \xi_{\cal I}(g=g_{\cal I}, L)} =
  {R_\xi(n, g=g_{\cal I},
    \kappa, \delta, L)\over  R_\xi(n,g=g_{\cal I},\kappa=0,\delta,L)}\,,
  \label{ratioconv}
\end{equation}
being $\xi_{\cal I}$ the correlation length of the standard Ising ring
at the QCP, are provided in the bottom panel of Fig.~\ref{figQscaling}
[note that we prefer to show the ratio~\eqref{ratioconv}, since
  it is subject to smaller scaling corrections than
  $R_\xi=\xi(n, g=g_{\cal I}, \kappa, \delta, L)/L$].

We point out that the above scaling behaviors are expected to be
approached with power-law suppressed corrections, which may depend on
the observable considered, analogously to the FSS of closed Ising
rings~\cite{CPV-15,RV-21,FRV-22}.  The scaling corrections
characterizing the approach to the asymptotic FSS are analogous to
those of Ising rings in the presence of symmetry-breaking defects
discussed in Ref.~\cite{FRV-22}.  We expect that such corrections
decay as $L^{-3/4}$ in the case of the ratios $R_\xi=\xi/L$ and
$\xi/\xi_{\cal I}$, since they arise from analytical
backgrounds~\cite{CPV-14,RV-21}. They are instead suppressed as
$L^{-7/4}$ in the case of the decoherence factor $Q$ and the gap ratio
$\Delta/\Delta_{\cal I}$. Such $O(L^{-7/4})=O(L^{-2y_\kappa})$ corrections
arise from the nonlinear scaling field~\cite{FRV-22} associated with
the Hamiltonian parameter $\kappa$, which is given by $u_\kappa =
\kappa + c \kappa^3 + \ldots$ (where the quadratic term vanishes, due to
the parity-symmetry properties of the Ising ring), and in particular
from its third-order term.  Plots to highlight the scaling corrections
for the observable considered are reported in the insets of
the three panels of Fig.~\ref{figQscaling} (notice the different
dependences on $L$ of the quantities in the $x$ axis, which match the
exponent of the various corrections). They are definitely compatible
with the expected finite-size power-law suppressions.

\begin{figure}[!t]
  \includegraphics[width=0.95\columnwidth]{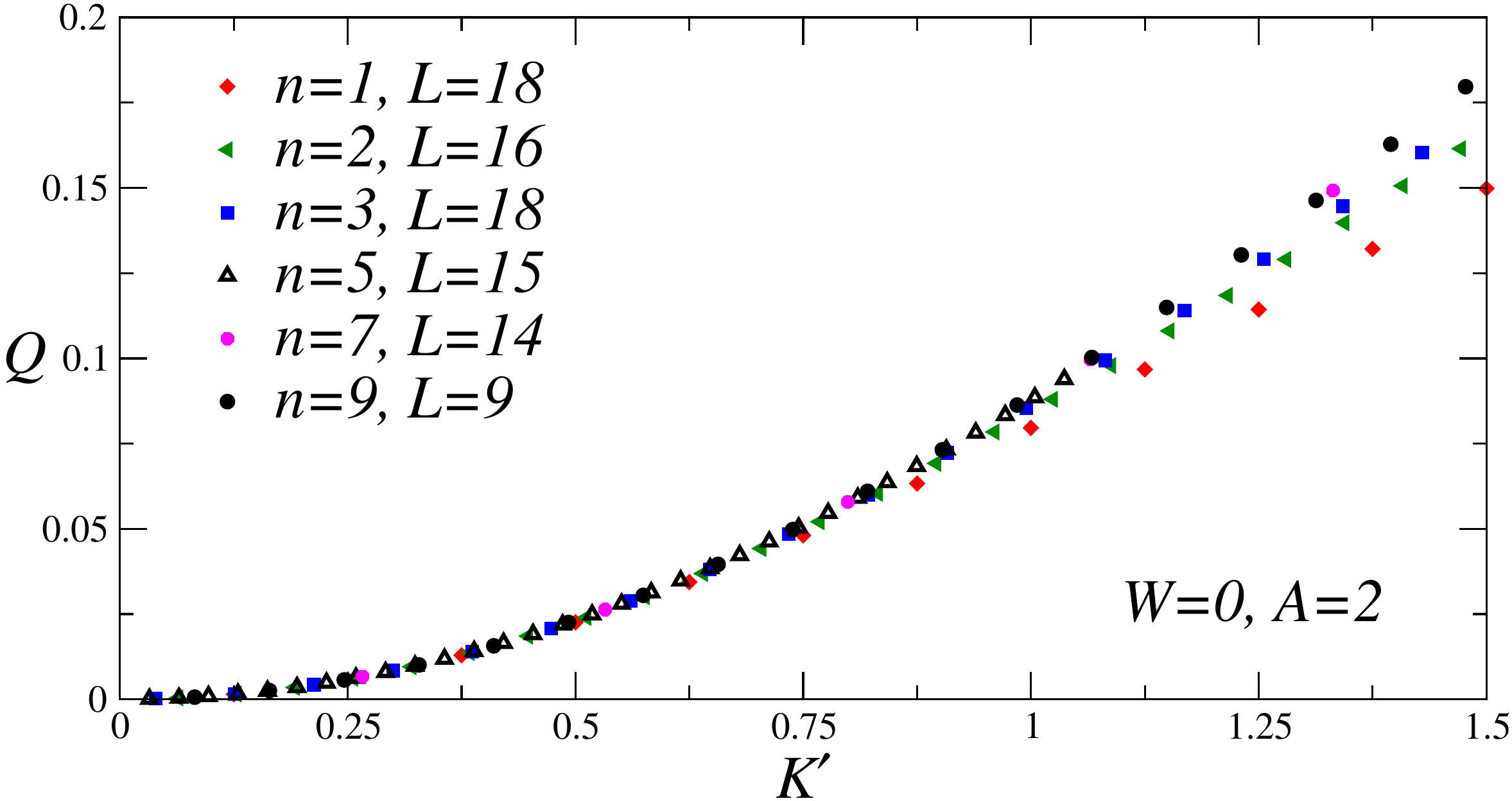}
  \includegraphics[width=0.95\columnwidth]{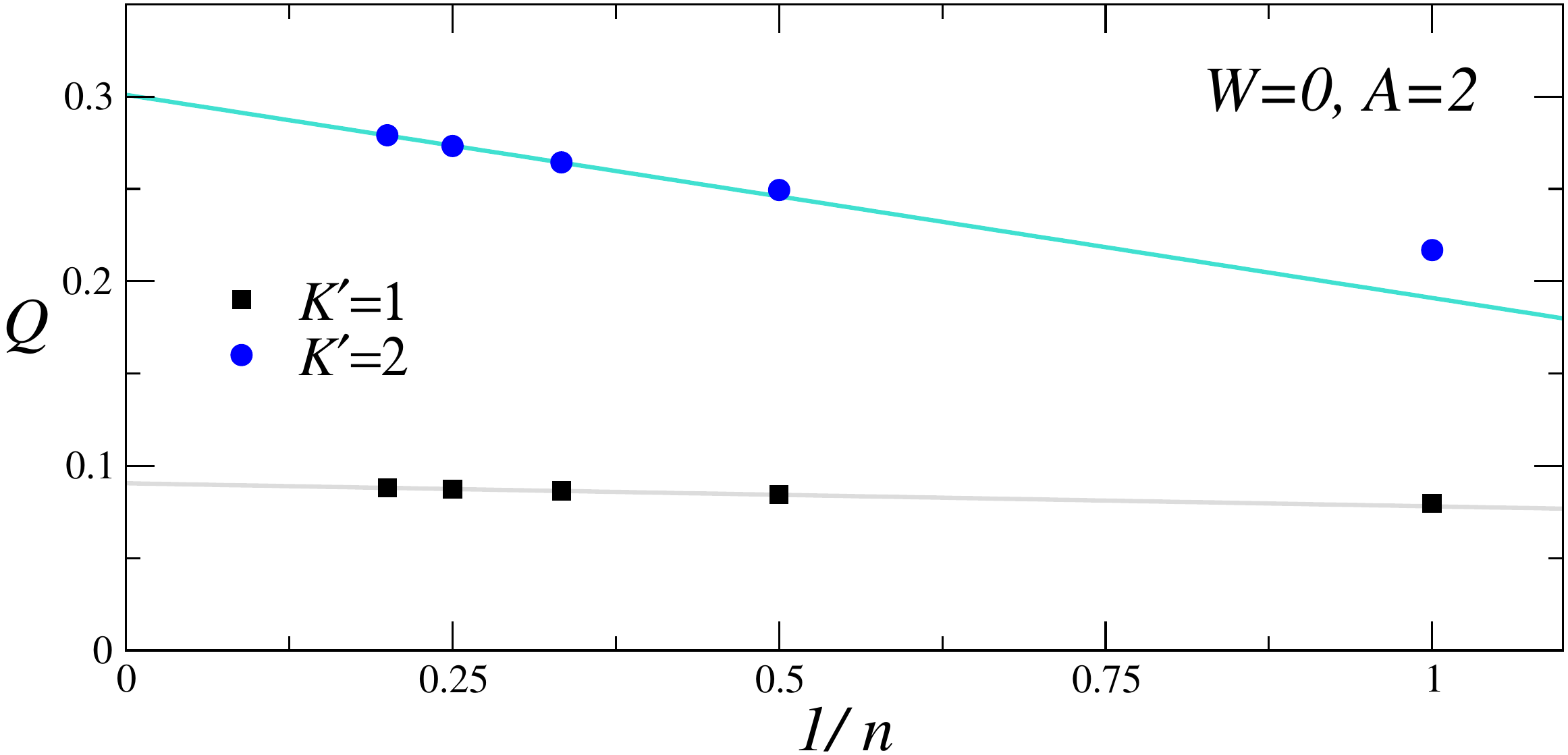}
  \caption{Top panel: decoherence factor $Q$ versus $K'$ at fixed
    $W=0, A=2$, and for several values of $n$. For each $n$, we only
    provide data for one of the largest lattice sizes at our
    disposal. However, we emphasize that scaling corrections are
    barely visible under the increase of $L$
    (cf.~Fig.~\ref{figQscaling}).  Bottom panel: $Q$ versus $1/n$ for
    fixed $K^\prime=1, 2$. Data are obtained by means of
    infinite-volume extrapolations ($L\to\infty$, at fixed $n$),
    assuming $L^{-7/4}$ scaling corrections. Straight lines are drawn
    to guide the eye. Extrapolations up to $n=5$ are compatible with a
    $1/n$ approach. The data provide a robust check of the scaling
    behavior reported in Eq.~\eqref{purityscan}.  }
  \label{figQmanyn}
\end{figure}

Our numerical results also suggest that the dependence on $n$ for a
sufficiently large number of ancillary qubits can be taken into
account through a redefinition of the scaling variables, after
introducing the variable
\begin{equation}
  K^\prime = \sqrt{n} \, K\,.
   \label{Kpdef}
\end{equation}
We provide evidence of this fact in Fig.~\ref{figQmanyn}, where the
decoherence factor $Q$ is shown versus $K^\prime$, for several values
of $n$ (top panel).  The plot supports the scaling behavior
\begin{equation}
  Q(n,g,\kappa,\delta,L) \approx \widetilde{\cal Q}(W,K^\prime,A)\,,
  \label{purityscan}
\end{equation}
with corrections that are apparently suppressed as $1/n$, as shown in
the bottom panel.  Analogous behaviors are observed for the other
quantities.  Note that the square-root power of $n$ in
Eq.~\eqref{Kpdef} is probably related to the fact that it actually
takes into account the effect of $n$ independent ancillary systems.

%%%%%%%%%%%%%%%%%%%%%%%%%%%%%%%%%%%%%%%%%%%%%%%%%%%%%%%%%%%%%%%%%%%%
%%%%%%%%%%%%%%%%%%%%%%%%%%%%%%%%%%%%%%%%%%%%%%%%%%%%%%%%%%%%%%%%%%%%

The above scaling behaviors show that the low-energy critical behavior
of the Ising ring experiences rapid and drastic changes under the
effects of perturbations arising from interactions with the ancillary
qubits, even in the case of one single ancillary isolated qubit.

\subsection{Behavior at finite ancillary energy scale $\delta$}
\label{finitekl}

As already mentioned, in the case of a finite number of ancillary
spins, the system remains critical at $g=g_{\cal I}$, even for finite value
of the energy scale $\delta$ and the interaction parameter $\kappa$.
The FSS behavior changes, similarly to changes arising from variations
of boundary conditions in closed critical systems. However the
variations for finite $\kappa$ and $\delta$ appear smooth, i.e., much
smoother than the drastic and rapid changes arising from their turning
on at small $\kappa$ and $\delta$ (see Sec.~\ref{smallkl}).

\begin{figure}[!t]
    \includegraphics[width=0.95\columnwidth]{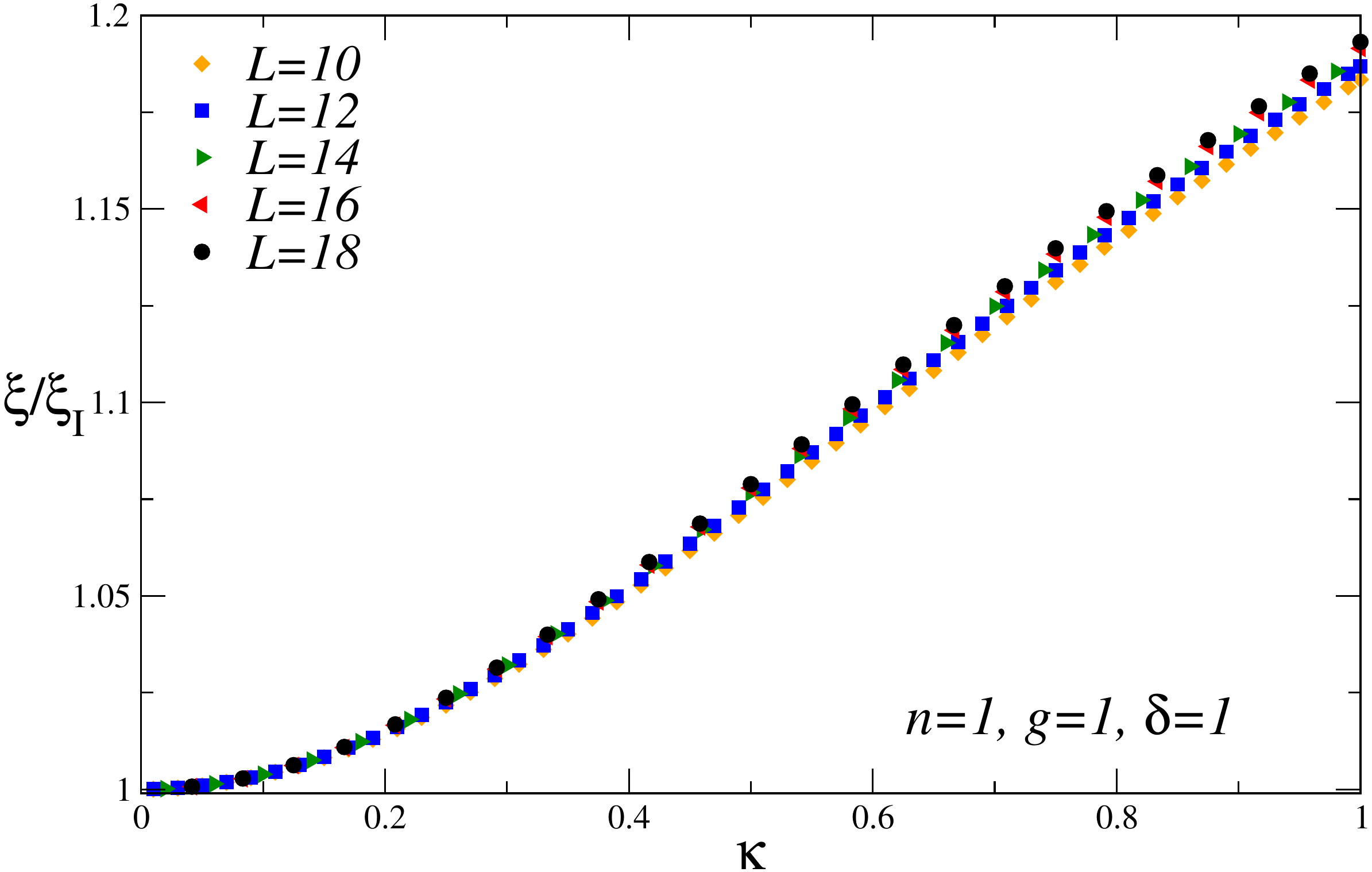}
    \includegraphics[width=0.95\columnwidth]{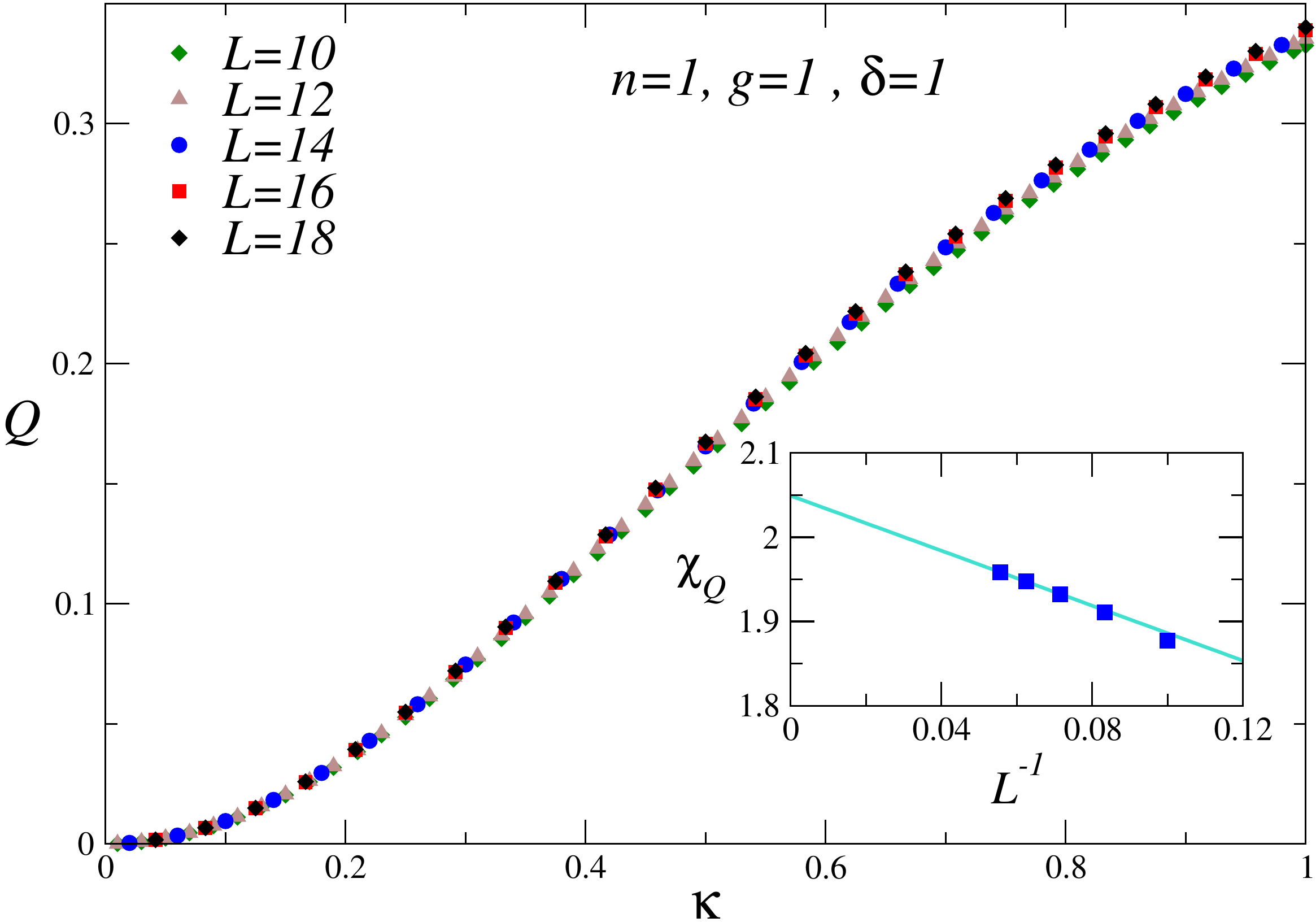}
    \caption{Ratio $\xi/\xi_{\cal I}$ (top) and decoherence factor $Q$
      (bottom) versus $\kappa$, for finite $n=1$, $\delta=1$, and
      $g=g_{\cal I}=1$.  The data show a smooth behavior with $\kappa$,
      without any significant dependence on the size of the Ising
      ring, suggesting that the effect at the QCP remains essentially
      local when $\delta>0$ is kept finite in the large-$L$ limit.
      The inset in the bottom panel shows that the decoherence
      susceptibility $\chi_Q$ converges to a constant as $L^{-1}$,
      in the large-volume limit.}
    \label{figQdeltafinite}
\end{figure}

In the following we consider the FSS limit, keeping the Hamiltonian
parameter $\delta>0$ fixed. The numerical results at $g=g_{\cal I}$ show that
the observables remains critical, but their FSS curves change when
varying $\kappa$ and $\delta$, without apparent nonanalyticities, like
those observed around $\kappa=\delta=0$ (top panel of
Fig.~\ref{figQdeltafinite}).
It is also worth noting that, when keeping $\delta>0$ fixed, the
decoherence factor $Q$ is expected to behave smoothly around
$\kappa=0$, thus its susceptibility $\chi_Q$ is not expected to
diverge in the large-$L$ limit, unlike around $\delta=0$.  This is
confirmed by the results in the bottom panel of
Fig.~\ref{figQdeltafinite}, where the smooth behavior of the
decoherence $Q$ and its susceptibility $\chi_Q$ completely agree with
the presented scenario.

%%%%%%%%%%%%%%%%%%%%%%%%%%%%%%%%%%%%%%%%%%%%%%%%%%%%%%%%%%%%%%%%%%%%
%%%%%%%%%%%%%%%%%%%%%%%%%%%%%%%%%%%%%%%%%%%%%%%%%%%%%%%%%%%%%%%%%%%%

\section{Sunburst Ising model at the CQT, with qubits at fixed distance}
\label{infiniten}

We now discuss a different situation, where the isolated qubits of the
ancillary system become infinite, and their contacts with the Ising
ring are maintained at a fixed distance $b$ [case (II) in
  Sec.~\ref{limits}].  Therefore, in the large-size limit, $L\to \infty$
and also $n = L/b \to \infty$.  Again we distinguish between two
regimes:\\ (i) the interaction with the ancillary qubits represents a
small perturbation. In that case, we describe the distortion of the
FSS behavior around the quantum transition of the Ising ring
(cf.~Sec.~\ref{ismallkl});\\ (ii) the effects of the interaction with
the ancillary qubits is substantial.  In that case, we show how the
phase diagram gets changed, still developing an Ising-like transition
at $g_c>g_{\cal I}$ (cf.~Sec.~\ref{ifinitekl}).

\subsection{FSS around the closed Ising-ring criticality}
\label{ismallkl}

As done in Sec.~\ref{smallkl}, in order to discuss the effects of the
perturbations arising from the ancillary spin system at the CQT of the
Ising ring, we should first determine the corresponding scaling
variables.  For this purpose, we make reasonable guesses based on the
results for a finite number of ancillary spins, and in particular for
a large number of $n$.  Our working hypothesis is that $A$, defined in
Eq.~\eqref{adef}, still remains a good scaling variable (essentially
because $\delta$ is again the gap of the ancillary spin system), but
we need to change $K$, cf. Eq.~\eqref{kdef}.  Indeed, using
Eq.~\eqref{Kpdef}, and noting that $n=L/b$ in the case at hand, we
arrive at the following expression:
\begin{equation}
  \widetilde{K} = \kappa \, b^{-1/2}\, L^{\tilde{y}_{\kappa}}\,,
  \qquad \tilde{y}_{\kappa}=y_\kappa + 1/2=11/8\,.
    \label{widetildeK}
\end{equation}

\begin{figure}
    \includegraphics[width=0.95\columnwidth]{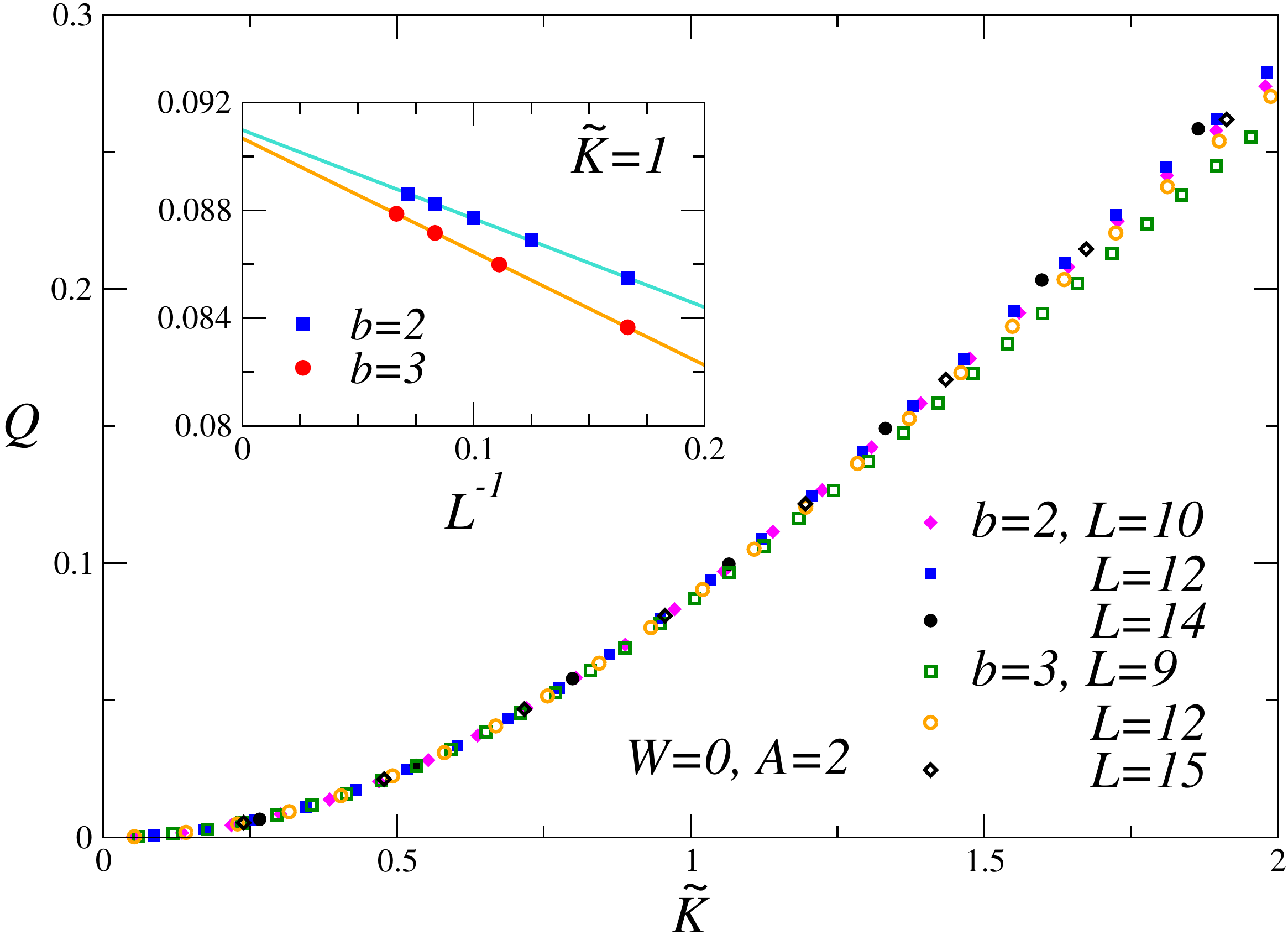}
    \caption{Scaling of the decoherence factor $Q$ versus
      $\widetilde{K}$, for fixed $W=0$, $A=2$, and $b=2, 3$. In the
      inset, large-volume extrapolations for $b=2$ and $b=3$ are
      consistent with a decay $L^{-1}$ and compatible with each
      other. These data definitely support the scaling behavior of
      Eq.~\eqref{qscanb}, thus implying the power-law divergence
      of the corresponding decoherence susceptibility: $\chi_Q \sim
      L^{2\tilde{y}_\kappa}$ with $2\tilde{y}_\kappa=11/4$, see
      Eq.~\eqref{qscanb2}. They also show that the parameter $b$ does
      not play a major role, apart from entering into the definition
      of $\widetilde{K}$.  }
    \label{figQscalingfixedb}
\end{figure}

Therefore, the decoherence properties around the Ising-ring
criticality are expected to obey the scaling behavior
\begin{equation}
  Q(b,g,\kappa,\delta,L) 
  \approx \widetilde{\cal Q}(W,\widetilde{K},A)\,.
  \label{qscanb}
\end{equation}
This implies that
\begin{equation}
  \chi_Q(b,g,\delta,L) 
  \approx L^{2\tilde{y}_\kappa} \widetilde{\cal C}(W,A)\,,
  \label{qscanb2}
\end{equation}
showing a faster divergence with respect to that for a finite number
of ancillary spins [compare with Eq.~\eqref{chiqsca}].  Numerical data
for the decoherence factor $Q$ reported in
Fig.~\ref{figQscalingfixedb} fully support this scaling ansatz, within
scaling corrections that, for fixed $b$, appear to be suppressed as
$L^{-1}$. Analogous behaviors are developed by the other observables
(not shown), for example in the case of RG-invariant quantities
\begin{equation}
  R(b,g,\kappa,\delta,L) 
  \approx \widetilde{\cal R}(W,\widetilde{K},A)\,.
  \label{Rscan}
\end{equation}
We also expect that the microscopic distance $b$ between the ancillary
contacts does not play a major role, apart from entering the
definition of the scaling variable $\widetilde{K}$.

%%%%%%%%%%%%%%%%%%%%%%%%%%%%%%%%%%%%%%%%%%%%%%%%%%%%%%%%%%%%%%%%%%%%
%%%%%%%%%%%%%%%%%%%%%%%%%%%%%%%%%%%%%%%%%%%%%%%%%%%%%%%%%%%%%%%%%%%%

\subsection{Phase diagram for finite $\kappa$ and $\delta$}
\label{ifinitekl}

\begin{figure}[!t]
    \includegraphics[width=0.95\columnwidth]{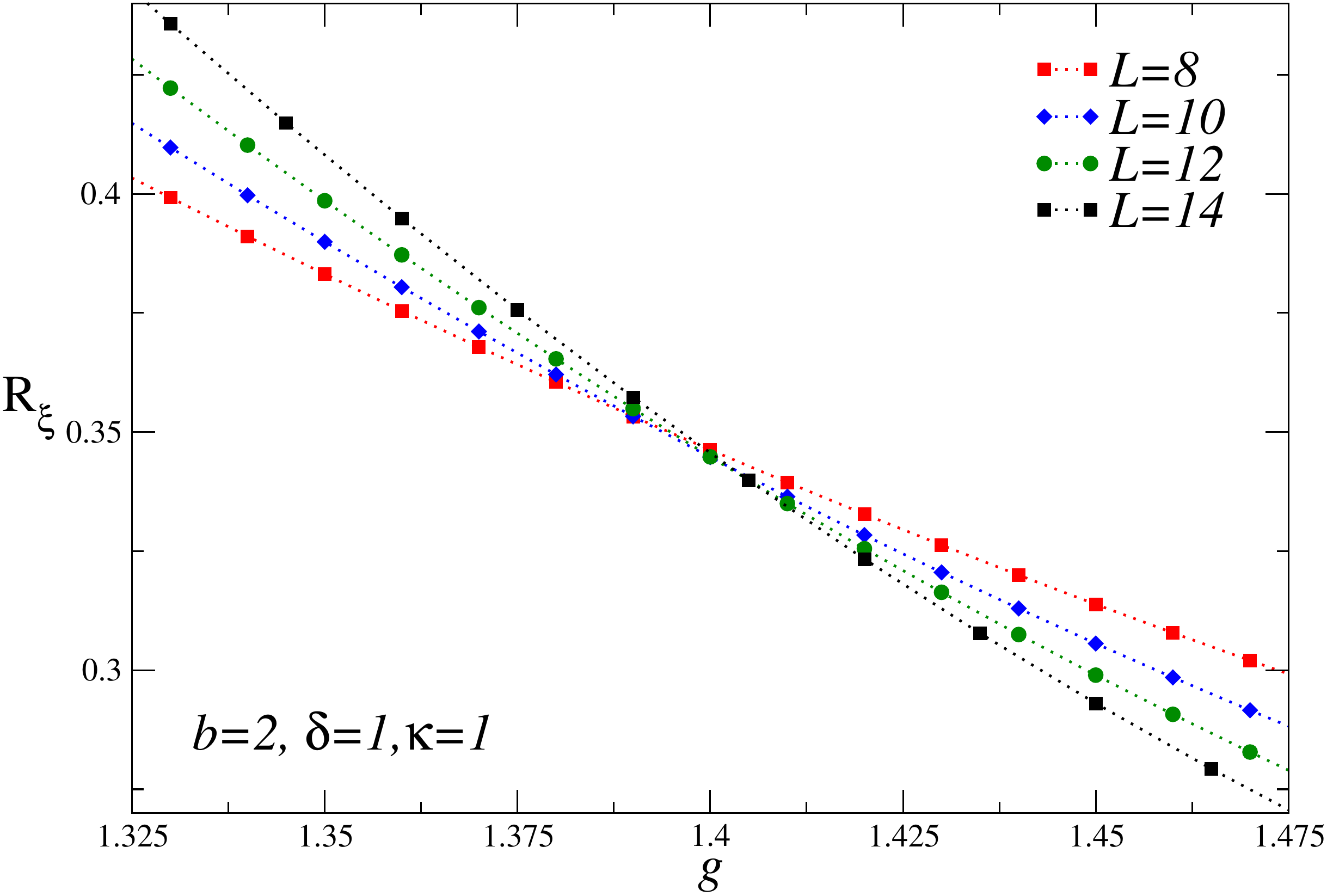}
    \caption{The RG-invariant quantity $R_\xi$ versus $g$, for $b=2$
      and finite $\delta=\kappa=1$. The crossing point $g_c(\kappa=1,
      \delta=1)\approx 1.40$ signals the Ising-like transition between
      the paramagnetic and the ordered phase of the lattice model,
      which moves away from the Ising critical point $g_{\cal I}=1$.}
    \label{figrxicrossingpoint2b}
\end{figure}

For finite values of the Hamiltonian parameters $\kappa$ and $\delta$,
the ancillary system may give rise to a modification of the phase
diagram.  In particular, we expect that the system at $g=g_{\cal I}$
does not remain {\em critical}.
Since the whole system has a global ${\mathbb Z}_2$ symmetry, we may
still have Ising-like transitions, but their positions could shift to
other values of $g$. This is clearly demonstrated by the numerical
results shown in Fig.~\ref{figrxicrossingpoint2b}, which provide
evidence of a transition along a surface at $g_c(\kappa,\delta)>g_{\cal I}$
(see Sec.~\ref{disbeh} for further details).

\begin{figure}[!t]
    \includegraphics[width=0.95\columnwidth]{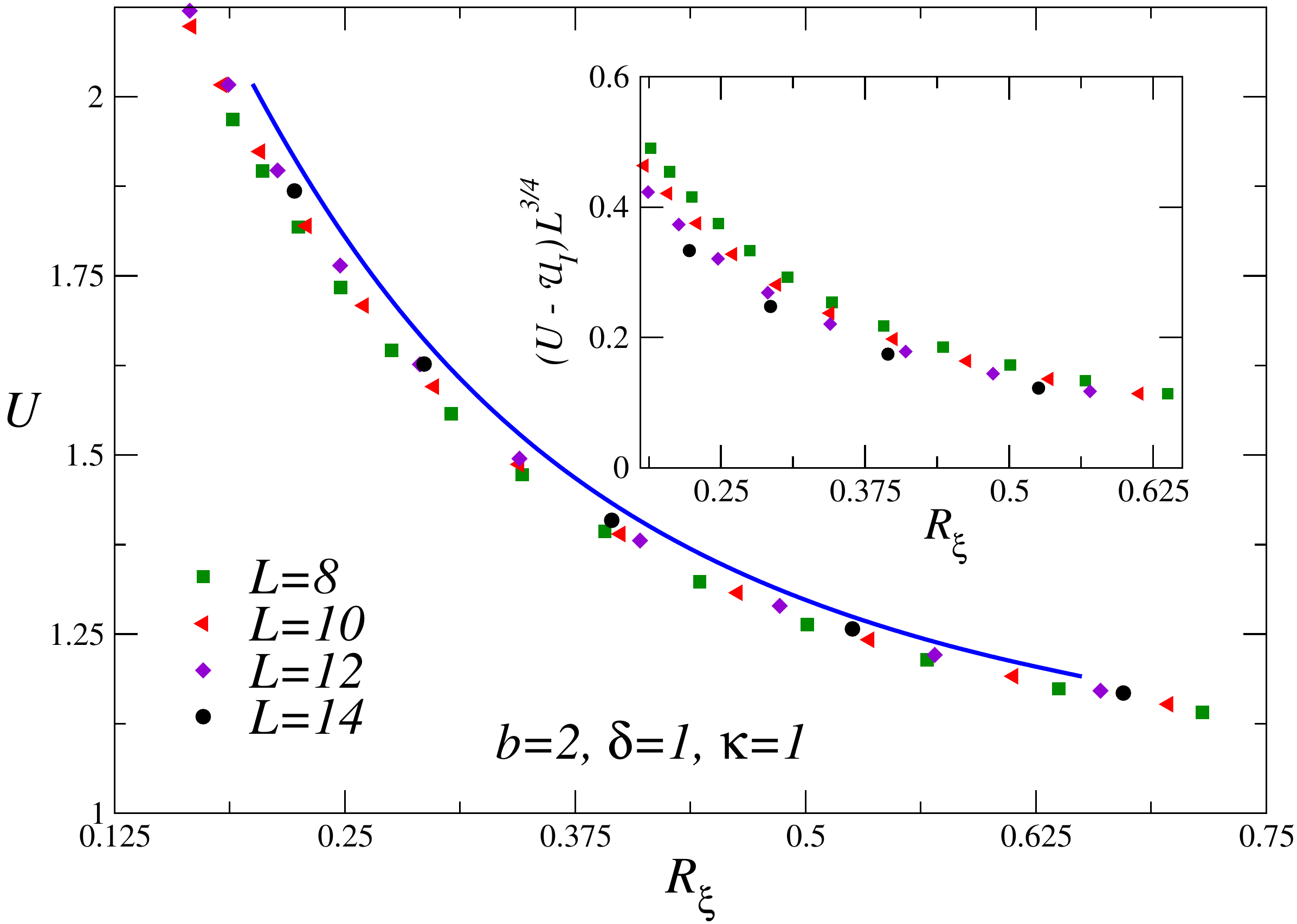}
    \caption{Binder parameter $U$ versus the RG-invariant ratio $R_\xi$,
      for $b=2$, $\delta=1$, and $\kappa=1$. The blue line represents
      the infinite-volume extrapolation of the universal curve
      $\mathcal{U}_{\cal I}(R_\xi)$ for the conventional Ising ring. Data for
      finite $\delta$ and $\kappa>0$ approach the same RG universal
      curve of the Ising universality class with PBC.  In the inset,
      we show $(U - \mathcal{U}_{\cal I})L^{3/4}$ in terms of $R_\xi$.
      The behavior of the scaling corrections appear substantially
      consistent with their expected $O(L^{-3/4})$ asymptotic suppression.}
    \label{figurxi2b}
\end{figure}

To address the nature of the transition, assuming that $R_\xi$ is a
monotonic function of $g$, one can invert Eq.~\eqref{rsca} and substitute
it into the Binder dependence, such that in the large-volume limit
\begin{equation}
  U(b, L, g, \delta, \kappa) \approx \mathcal{U}_{\cal I}(R_\xi)\,.
\end{equation}
This relation is particularly useful to compare the universality class
of different models, as the function $\mathcal{U}_{\cal I}$ only depends on
the boundary conditions and the universality class at the transition.
Moreover, it does not require the tuning of any
non-universal parameter to be satisfied, so the universality class of
two different lattice models can be easily compared. Leading scaling
corrections to this formula are expected to get suppressed as
$L^{-3/4}$~\cite{CPV-14,RV-21}.  In Fig.~\ref{figurxi2b} we show
results for the Binder parameter $U$ vs.~the RG-invariant quantity
$R_\xi=\xi/L$ [cf. Eqs.~\eqref{Udef} and~\eqref{xidef}, respectively]
at finite values of $\delta$ and $\kappa$, together with the
interpolation of the universal Ising curve $\mathcal{U}_{\cal I}(R_\xi)$
for PBC.

%%%%%%%%%%%%%%%%%%%%%%%%%%%%%%%%%%%%%%%%%%%%%%%%%%%%%%%%%%%%%%%%%%%%
%%%%%%%%%%%%%%%%%%%%%%%%%%%%%%%%%%%%%%%%%%%%%%%%%%%%%%%%%%%%%%%%%%%%

%%%%%%%%%%%%%%%%%%%%%%%%%%%%%%%%%%%%%%%%%%%%%%%%%%%%%%%%%%%%%%%%%%%%
%%%%%%%%%%%%%%%%%%%%%%%%%%%%%%%%%%%%%%%%%%%%%%%%%%%%%%%%%%%%%%%%%%%%

To better understand the features of the phase diagram in the $(g,
\kappa)$ plane, we keep $\delta > 0$ fixed and discuss the behavior
when varying $\kappa$ around $\kappa=0$, around the pure Ising-ring
QCP ($g=g_{\cal I}$). Note that this is different from the case
considered in Sec.~\ref{ismallkl}, where $\delta$ was rescaled as
$\delta \sim L^{-z}$. Therefore the scaling behaviors outlined in
Sec.~\ref{ismallkl} are not expected to hold anymore.

To describe how the QCP moves, we may analyze the behavior of the
scaling field $u_g$ under the effect of $\kappa$. While in the
previous analyses we considered $u_g=g-g_{\cal I}$, cf.~Eq.~\eqref{wdef},
here we should take into account the presence of the interaction with
the ancillary system, and therefore its dependence on $\kappa$. Since
the global system is invariant when changing the sign of $\kappa$,
then $u_g$ must be an even function of $\kappa$.  The additional
requirement of analyticity with respect to the Hamiltonian parameters
then gives:
\begin{equation}
  u_g \approx g- g_{\cal I} - C_\kappa(\delta,b) \kappa^2\,.
  \label{ug}
\end{equation}
Note that no terms containing only $\delta$ must be present, since we
must recover the closed Ising-ring expression when $\kappa=0$ for any
$\delta>0$.

%%%%%%%%%%%%%%%%%%%%%%%%%%%%%%%%%%%%%%%%%%%%%%%%%%%%%%%%%%%%%%%%%%%%
\begin{figure}[!t]
  \includegraphics[scale=0.35]{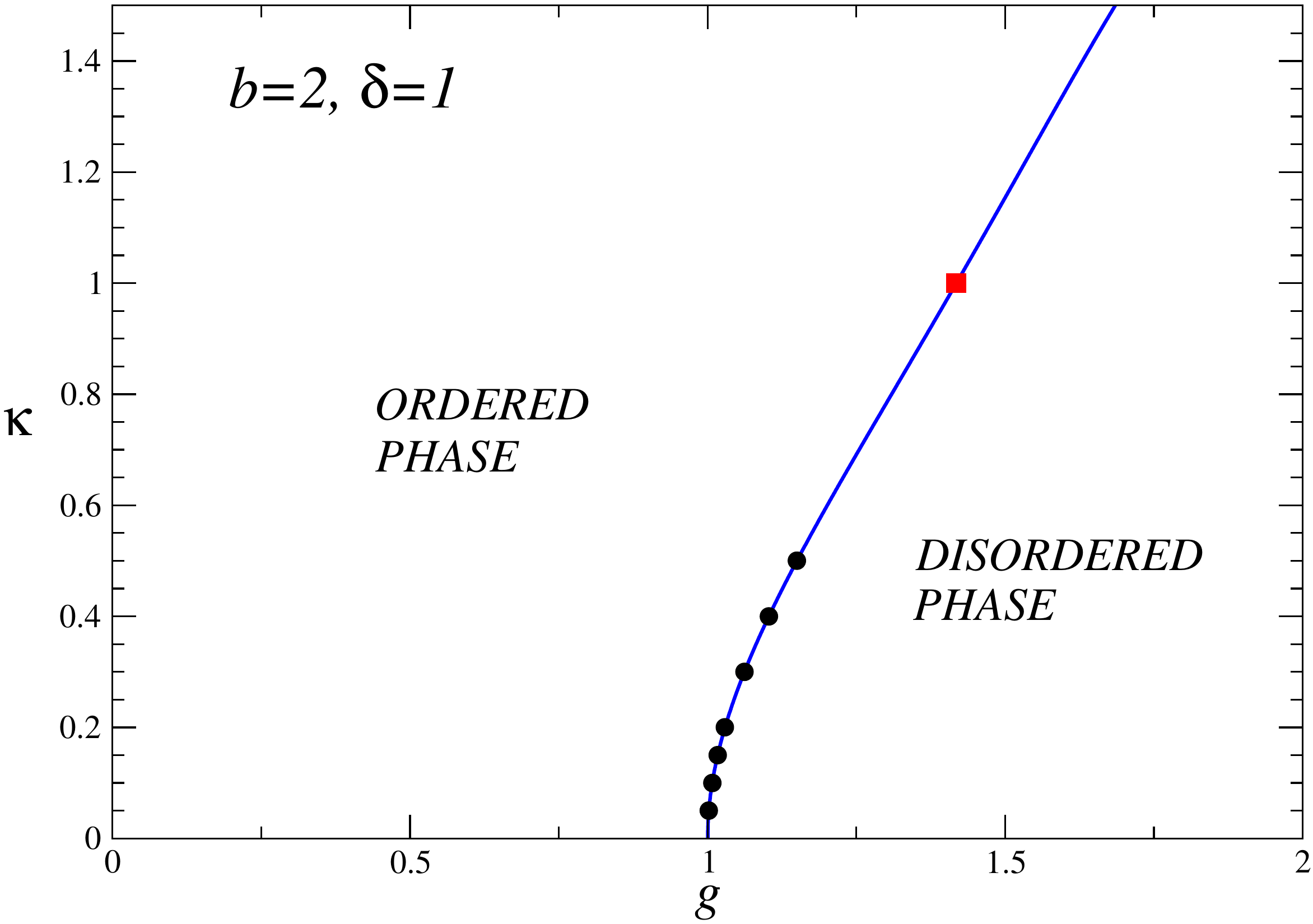} \caption{A
      sketch of the $g$-$\kappa$ phase diagram of the sunburst quantum
      Ising model with $b=2$, at $\delta=1$.  The red square at
      $\kappa=1$ corresponds to the specific case analyzed in
      Fig.~\ref{figrxicrossingpoint2b}, while the black circles have
      been obtained by analyzing the RG-invariant quantity $\xi /
      \xi_I$ [cf.~Eq.~\eqref{ratioconv}], up to $L=12$.  The estimates
      of the critical points are fully consistent with the behavior
      reported in Eq.~(\ref{gcbeh}), i.e.  $g_c(\kappa)-g_{{\cal I}}
      \sim \kappa^2$ for sufficiently small values of $\kappa$.}
  \label{fig:phasediagram}
\end{figure}
%%%%%%%%%%%%%%%%%%%%%%%%%%%%%%%%%%%%%%%%%%%%%%%%%%%%%%%%%%%%%%%%%%%%

The critical point occurs when $u_g=0$, thus we expect 
\begin{equation}
  g_c(\kappa) \approx g_{\cal I} + C_\kappa(\delta,b) \kappa^2\,,
  \label{gcbeh}
\end{equation}
which should hold for sufficiently small values of $\kappa$.
This behavior has been numerically checked by estimating the
critical points $g_c(\kappa)$ for various values of $\kappa$, see
Fig.~\ref{fig:phasediagram}.

\begin{figure}
  \includegraphics[width=0.95\columnwidth]{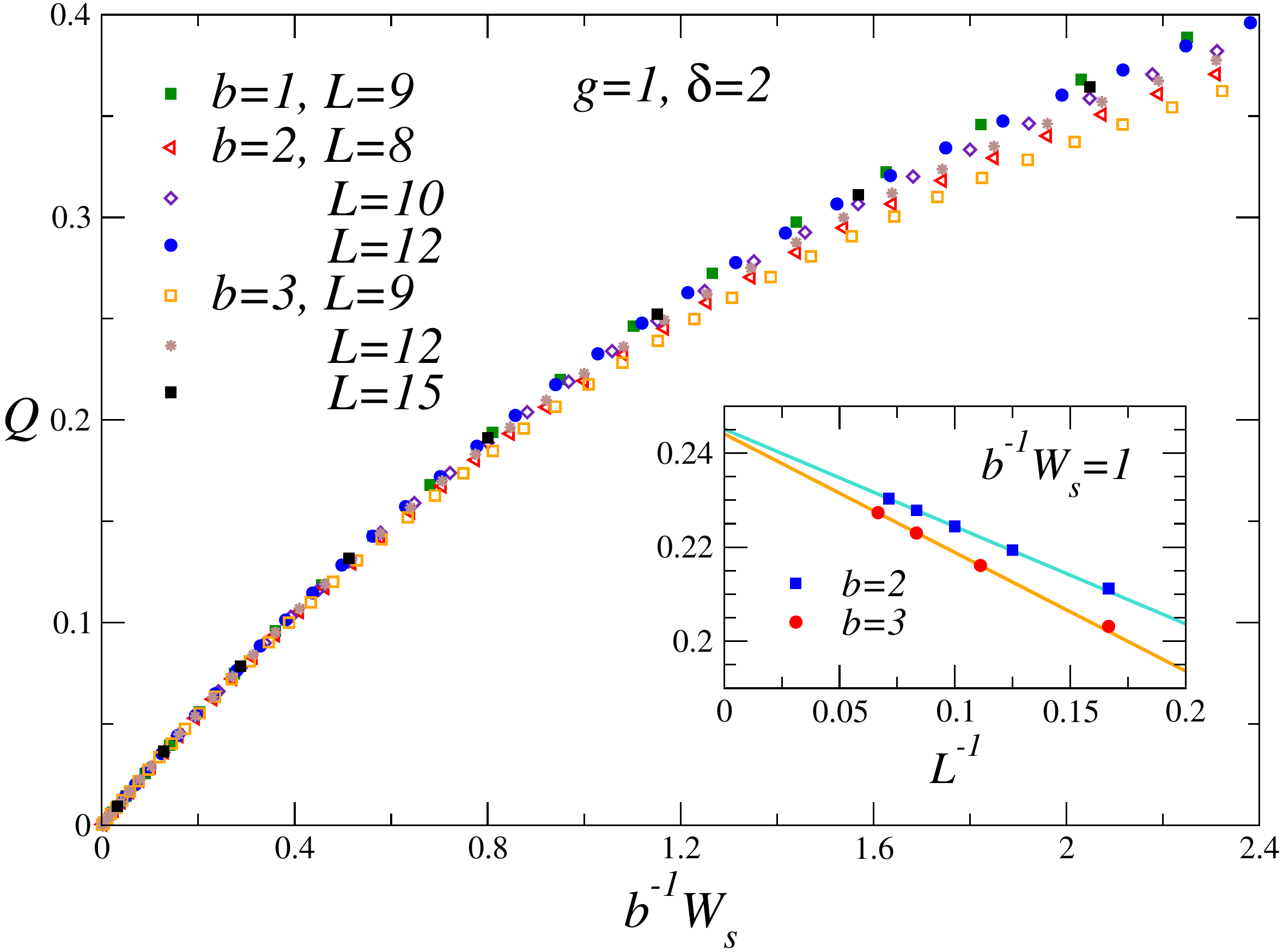}
  \caption{$Q$ versus $b^{-1}\,W_s$ for finite $g=g_{\cal I}, \delta=2$ and
    various values of $b$. The data nicely confirm the FSS
    Eq.~\eqref{Qscangi}.  In the inset, scaling corrections at $b^{-1}
    W_s=1$ are compatible with a $L^{-1}$ behavior. Note that the
    large-volume extrapolations for the two shown values of $b=2$ and
    $b=3$ are compatible with each other.  The straight lines are
    drawn to guide the eye.}
  \label{figdecscalingdeltafinitemanyb}
\end{figure}

Therefore, assuming Ising transitions related to the breaking of the
global ${\mathbb Z}_2$ symmetry, we expect to observe a scaling
behavior around $g_c$ in terms of the scaling variable
\begin{equation}
  \widetilde{W} = (g-g_c) L^{y_g}\,, \qquad y_g = 1\,.
  \label{tildew}
\end{equation}
This also implies, at $g=g_{\cal I}$ and for small $\kappa$, the scaling
behavior (keeping $\delta>0$ fixed)
\begin{eqnarray}
  Q(g=g_{\cal I},b,\kappa,\delta,L) & \approx & Q_s(\delta,W_s)\,,
  \label{Qscangi}\\
  R(g=g_{\cal I},b,\kappa,\delta,L) & \approx & R_s(\delta,W_s)\,,
  \label{Rscangi}
\end{eqnarray}
where
\begin{equation}
  W_s = \kappa^2 L \sim \widetilde{W}(g=g_{\cal I})\,.
  \label{widehatdef}
  \end{equation}
The above continuous transition line separates the phase diagram in
ordered and disordered phases, respectively for $g<g_c$ and $g>g_c$.
Results inherent to the presented FSS ansatz are shown in
Fig.~\ref{figdecscalingdeltafinitemanyb}, where the decoherence factor
$Q$ nicely scales in terms of $b^{-1}\,W_s$.  This plot also supports
the fact that the coefficient $C_\kappa$ of the $\kappa^2$ term in
Eq.~\eqref{gcbeh} behaves as $C_\kappa\approx b^{-1}f(\delta)$,
implying again that the microscopic distance $b$ does not play a major
role, apart from entering into the redefinition of the scaling
variable $W_s$.

%%%%%%%%%%%%%%%%%%%%%%%%%%%%%%%%%%%%%%%%%%%%%%%%%%%%%%%%%%%%%%%%%%%%
%%%%%%%%%%%%%%%%%%%%%%%%%%%%%%%%%%%%%%%%%%%%%%%%%%%%%%%%%%%%%%%%%%%%

On the basis of the above results, in Fig.~\ref{fig:phasediagram}
we provide a sketch of the phase diagram of the sunburst Ising model
with $b=2$, at $\delta=1$, in the space of parameters $g$ and $\kappa$.
An analogous scenario is expected to hold for any $\delta>0$
and any value of $b \ge 1$.

\section{Behavior along the FOQT line of the Ising ring}
\label{foqtbeh}

We now discuss the effects of the interaction with the ancillary
system at the FOQT line of the Ising ring, i.e., when $g<g_{\cal I}$,
and small values of $\kappa$ and $\delta$, around $\kappa=\delta=0$.
Again we address the problem within a FSS framework, which has been
recently developed also at FOQTs (see, e.g., Refs.~\cite{CNPV-14,RV-21}).
We consider the interaction with the ancillary system as a
perturbation, in particular when also its energy scale $\delta$ is
small.  We recall that, at a FOQT, the gap $\Delta_{\cal I}(g,L)$ is
generally exponentially suppressed in the case of neutral boundary
conditions~\cite{CNPV-14,PRV-18-fo}, such as for Ising rings,
cf. Eq.~\eqref{deltapbc}.

\subsection{The case of a finite number of ancillary spins}
\label{fnfoqt}

%%%%%%%%%%%%%%%%%%%%%%%%%%%%%%%%%%%%%%%%%%%%%%%%%%%%%%%%%%%%%%%%%%%%
\begin{figure}[!t]
    \includegraphics[width=0.95\columnwidth]{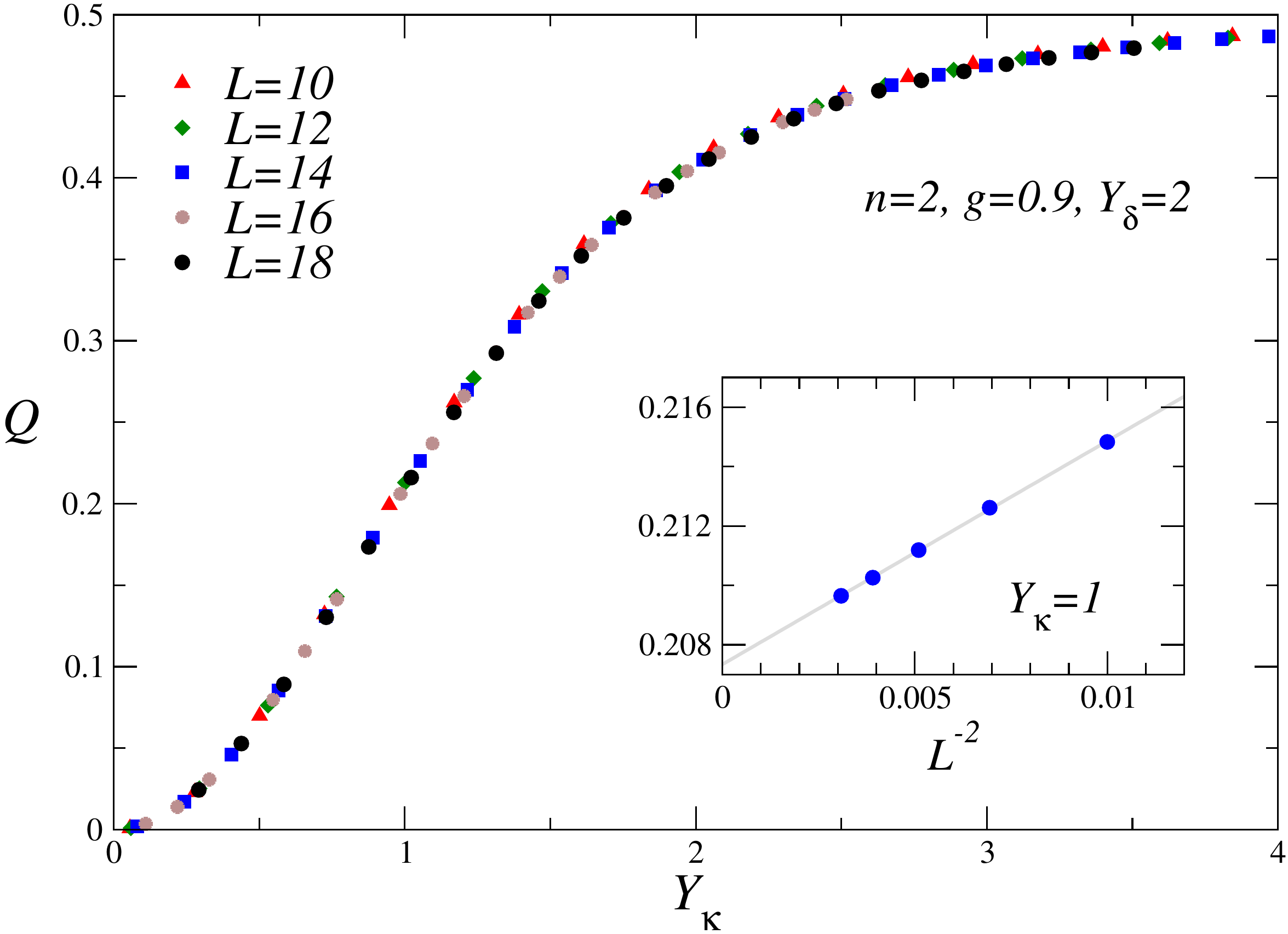}
    \caption{Scaling of the decoherence factor $Q$ versus $Y_\kappa$
      at a FOQT with $n=2, g=0.9$, and $Y_\delta=2$. Data nicely
      collapse with increasing the lattice size, supporting the
      presented FSS theory [see, in particular, Eq.~\eqref{purityscafo}].
      In the inset, scaling corrections at
      $Y_\kappa=1$ are in complete agreement with a $L^{-2}$
      decay. The straight line is drawn to guide the eye.}
    \label{figQscalingfirstorder2n}
\end{figure}
%%%%%%%%%%%%%%%%%%%%%%%%%%%%%%%%%%%%%%%%%%%%%%%%%%%%%%%%%%%%%%%%%%%%

\begin{figure}[!t]
\includegraphics[width=0.95\columnwidth]{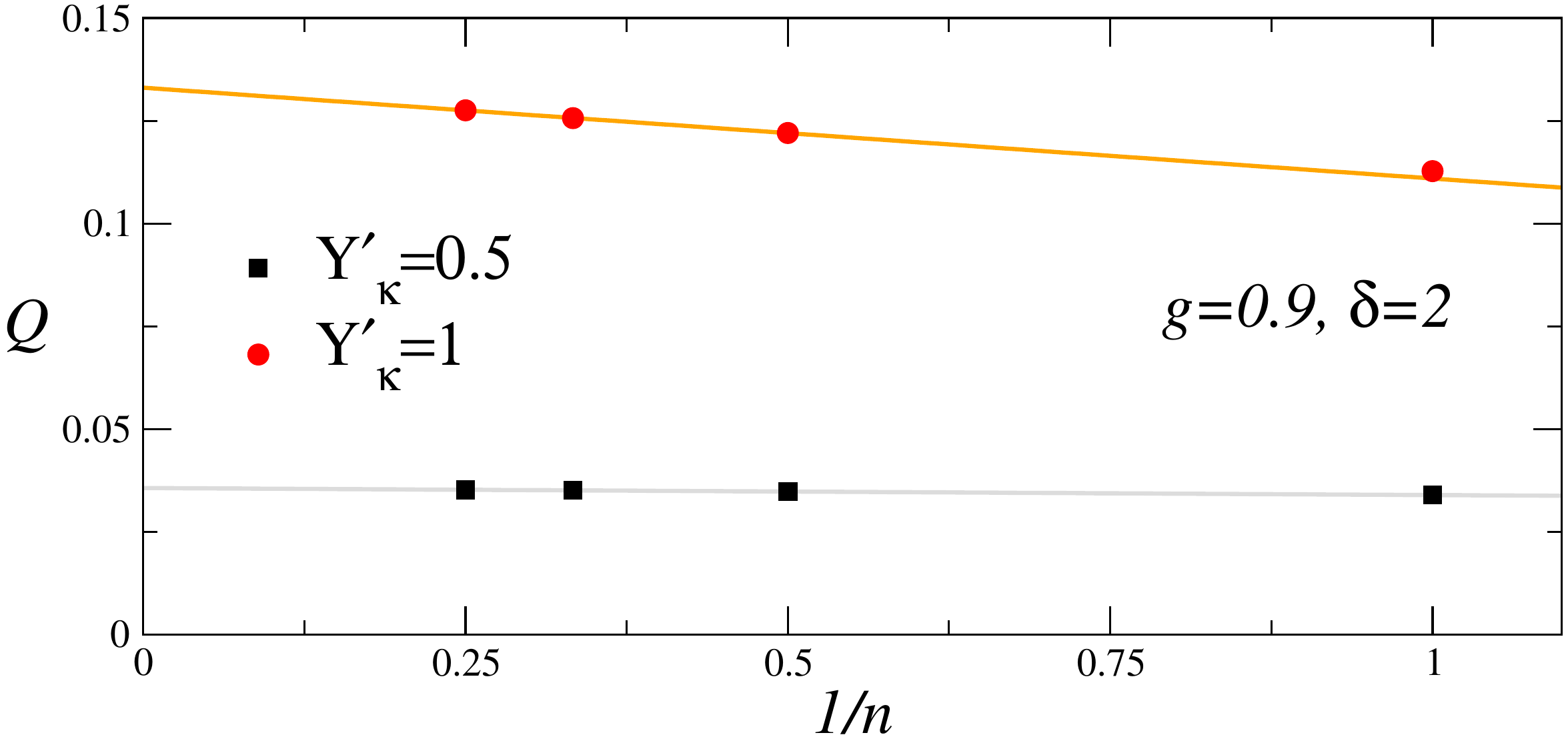}
\includegraphics[width=0.95\columnwidth]{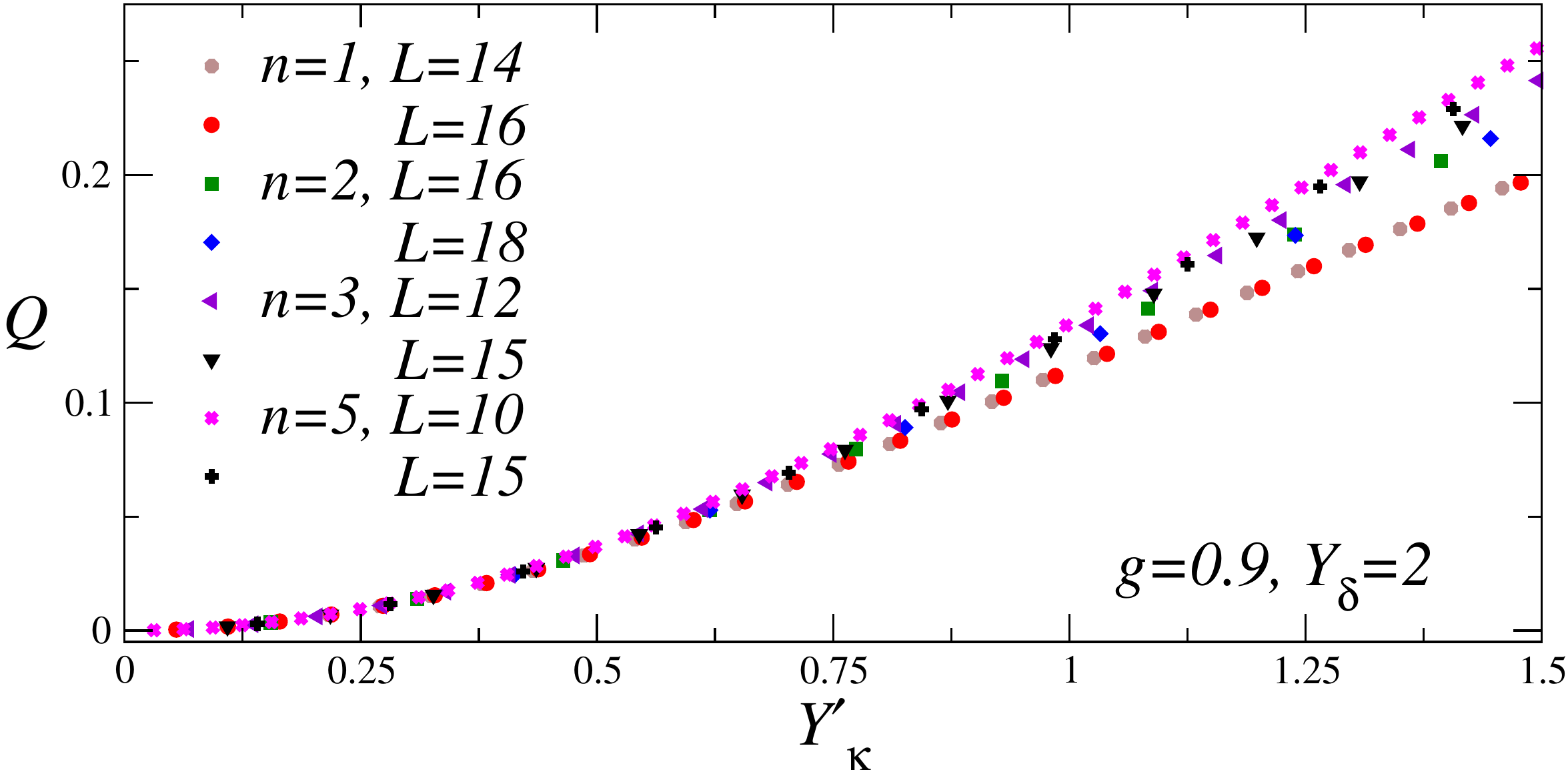}
    \caption{Top: Decoherence factor $Q$ versus the inverse number of
      ancillary spins $1/n$, for fixed $Y^\prime_\kappa=0.5$ and $1$.
      For each value of $n$, the infinite-volume extrapolations for
      the decoherence are computed assuming $L^{-2}$ scaling
      corrections.  Bottom: Decoherence factor $Q$ versus
      $Y^\prime_\kappa$ for many $n$ and $Y_\delta=2$ at a FOQT
      ($g=0.9)$. Data collapse improves along the whole curve with
      increasing the number of ancillary spins. These plots clearly
      support the scaling Eq.~\eqref{purityscafo2}, when $n$ is
      increased. Indeed the data appear to approach a scaling
      curve in the large-$n$ limit, which is clearly distinct
      from the curve for $n=1$, at least for sufficiently large
      values of $Y^\prime_\kappa$.}
    \label{figQfirstordermanyn}
\end{figure}

We follow the same ideas at the basis of the FSS description of FOQTs
of Ising systems driven by an external magnetic field (see, in
particular, Ref.~\cite{RV-21}).  We introduce the scaling variables
\begin{equation}
  Y_\kappa = {\kappa \over \Delta_{\cal I}(g,L)}\,,\qquad Y_\delta =
  {\delta \over \Delta_{\cal I}(g,L)}\,.
  \label{KAfoqt}
\end{equation}
Here $Y_\kappa$ is proportional to the ratio of the energy change arising
from the perturbation associated with $\kappa$ and the exponentially
suppressed gap of the pure Ising ring~\cite{CNPV-14}, while $Y_\delta$
is the ratio between the gap of the unperturbed ancillary system and
that of the Ising ring.  Therefore, we consider a FSS limit keeping
the scaling variables $Y_{\kappa}$ and $Y_{\delta}$ fixed.  The decoherence
factor defined in Eq.~\eqref{Qdef} is expected to behave as
\begin{equation}
  Q(n,g,\kappa,\delta,L) \approx {\cal Q}(n,Y_\kappa,Y_\delta)\,.
  \label{purityscafo}
\end{equation}
Numerical results for the scaling of $Q$ in
Fig.~\ref{figQscalingfirstorder2n} fully support the presented FSS
hypothesis.  Note that this also implies
\begin{equation}
  \chi_Q(n,g,\delta,L) \approx \Delta_{\cal I}(g,L)^{-2}{\cal C}(n,Y_\delta)\,.
  \label{chiqscafo}
\end{equation}
In turn, this implies that, since $\Delta_{\cal I}(g,L)$ is exponentially
suppressed, cf. Eq.~\eqref{deltapbc}, the coherence shows an exponentially
rapid drop around $\kappa=0$.
Like the behavior at the CQT, we again observe that the dependence on
$n$ can be absorbed within the scaling variables for sufficiently
large $n$, by replacing $Y_\kappa$ with
\begin{equation}
  Y_\kappa^\prime = \sqrt{n} \, Y_\kappa\,.
  \label{Ykappap}
\end{equation}
This clearly emerges from the plot in Fig.~\ref{figQfirstordermanyn},
where the decoherence factor $Q$ is shown as a function of
$Y_\kappa^\prime$ at fixed $Y_\delta$, for several values of $n$. The
analysis follows the same reasoning of the one carried out in the case
of CQTs (see Fig.~\ref{figQmanyn}).  Indeed, the data are consistent
with an asymptotic scaling behavior
\begin{equation}
  Q(n,g,\kappa,\delta,L) \approx {\cal Q}(Y_\kappa^\prime,Y_\delta)
  \label{purityscafo2}
\end{equation}
in the FSS limit, keeping $Y_\kappa^\prime$ fixed.

\begin{figure}
    \includegraphics[width=0.95\columnwidth]{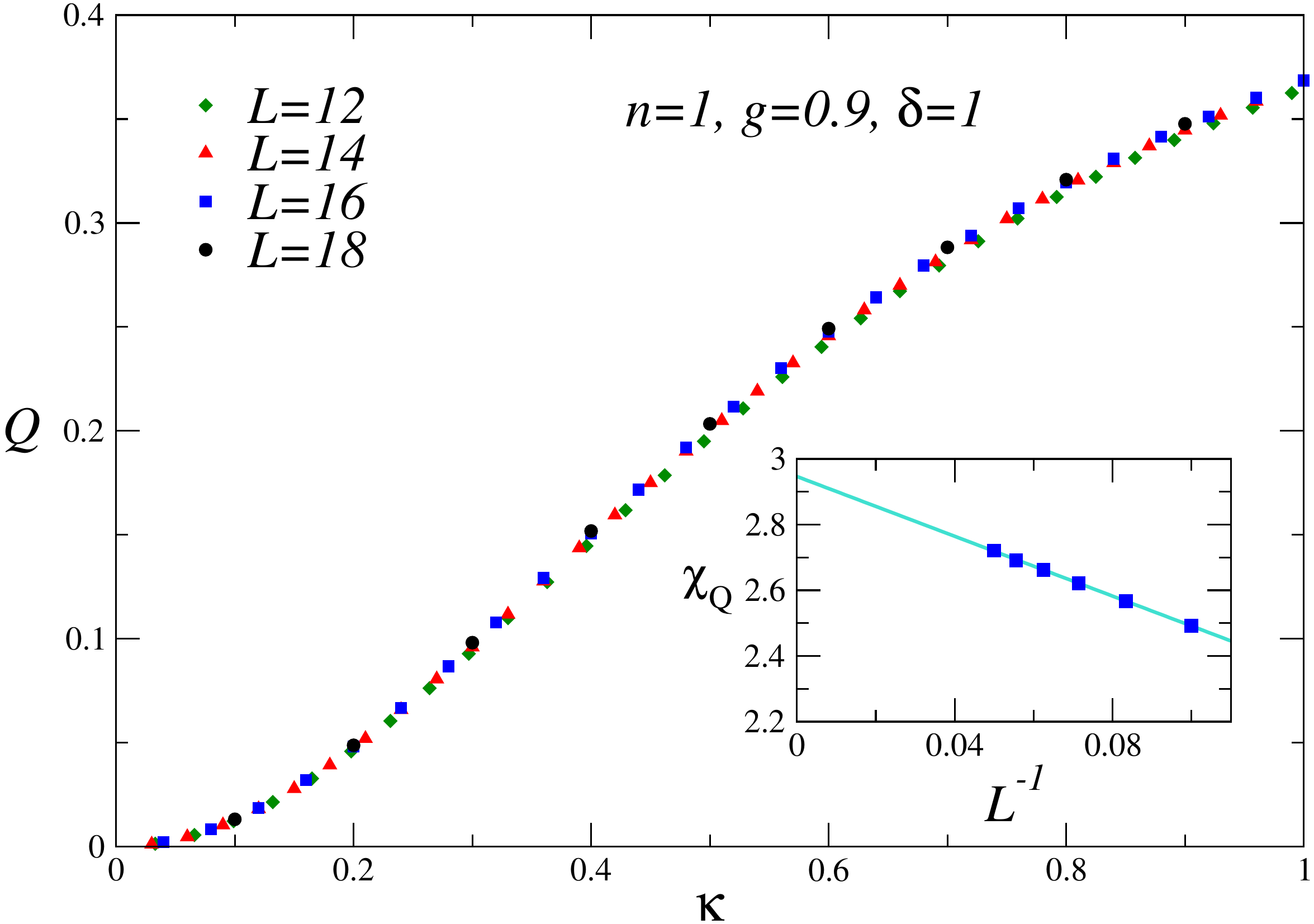}
    \caption{Decoherence factor $Q$ versus $\kappa$ for finite
      $g=0.9$, $\delta=1$, with a single ancillary spin. The plot shows
      that $Q$ behaves as a smooth function of $\kappa$, as well as
      its susceptibility $\chi_Q$, which converges to a constant
      as $L^{-1}$, in the large-volume limit (see the inset,
      where the line is drawn to guide the eye).}
    \label{figQfirstorderfinitedelta}
\end{figure}

Analogously to the behavior observed at the CQT, the above scaling
behavior changes significantly when $\delta>0$ is kept fixed in the
large-size limit.  In this case, the decoherence factor appears to
depend smoothly on $\kappa$, even at $\kappa=0$.  Therefore, also the
corresponding susceptibility $\chi_Q$ does not diverge in the
large-size limit.  Results are shown in
Fig.~\ref{figQfirstorderfinitedelta} for $\delta=1$.

%%%%%%%%%%%%%%%%%%%%%%%%%%%%%%%%%%%%%%%%%%%%%%%%%%%%%%%%%%%%%%%%%%%%
%%%%%%%%%%%%%%%%%%%%%%%%%%%%%%%%%%%%%%%%%%%%%%%%%%%%%%%%%%%%%%%%%%%%

\subsection{The case of ancillary spins at fixed distance}
\label{fnfoqtfd}

%%%%%%%%%%%%%%%%%%%%%%%%%%%%%%%%%%%%%%%%%%%%%%%%%%%%%%%%%%%%%%%%%%%%
%%%%%%%%%%%%%%%%%%%%%%%%%%%%%%%%%%%%%%%%%%%%%%%%%%%%%%%%%%%%%%%%%%%%

%%%%%%%%%%%%%%%%%%%%%%%%%%%%%%%%%%%%%%%%%%%%%%%%%%%%%%%%%%%%%%%%%%%%
\begin{figure}[!t]
    \includegraphics[width=0.95\columnwidth]{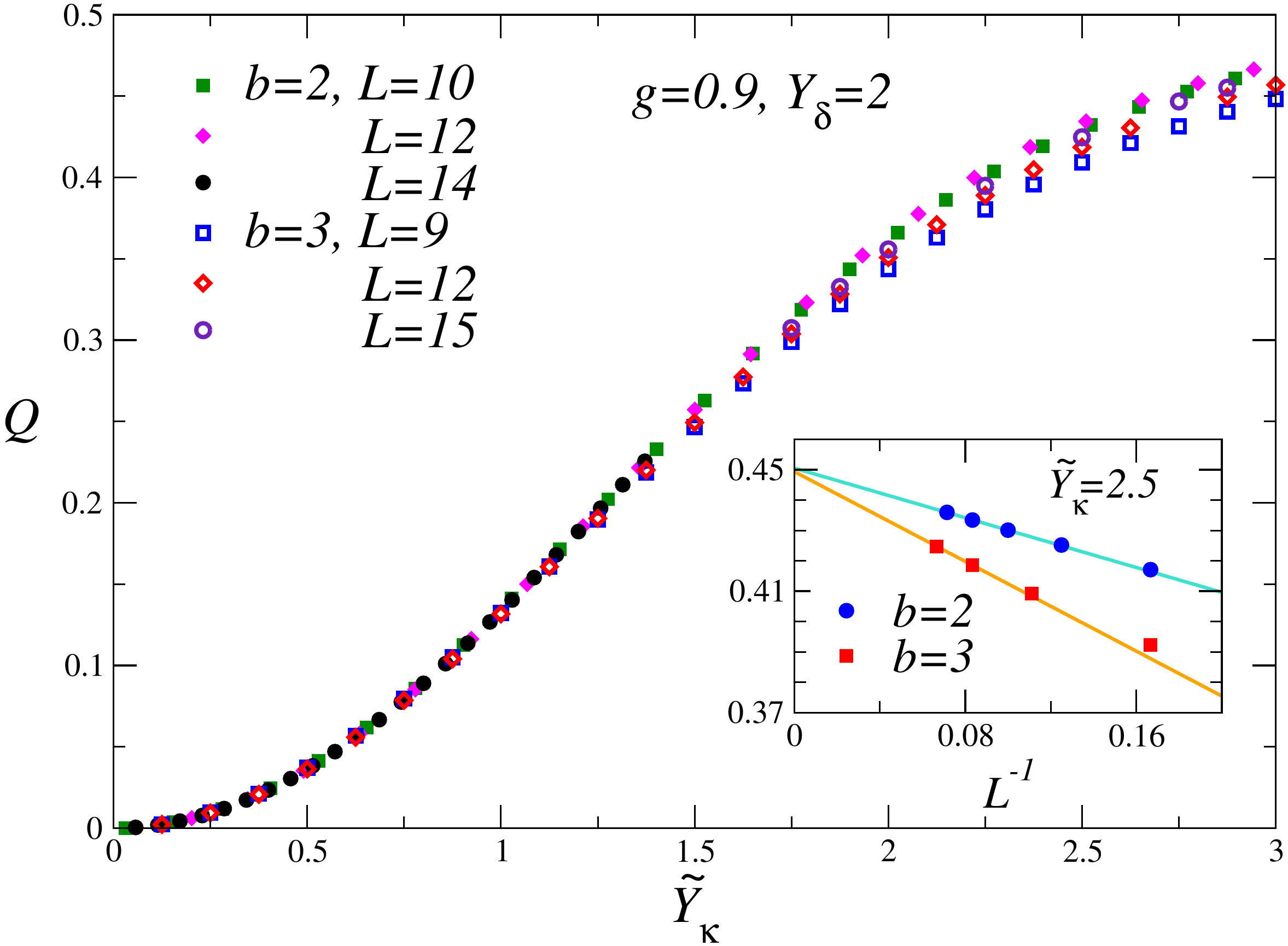}
    \caption{Decoherence factor $Q$ versus $\widetilde{Y}_\kappa$ for
      $Y_\delta=2$ and $b=2$ and $3$, at a FOQT ($g=0.9$). Data nicely
      collapse with increasing the lattice size, in agreement with the
      scaling behavior reported in Eq.~\eqref{purityscafo3}.  In the
      inset, scaling corrections at fixed $\widetilde{Y}_\kappa=2.5$
      appear to decay as $L^{-1}$.}
    \label{figQscalingfirstorder2b}
\end{figure}
%%%%%%%%%%%%%%%%%%%%%%%%%%%%%%%%%%%%%%%%%%%%%%%%%%%%%%%%%%%%%%%%%%%%

We now discuss the behavior at the FOQT of the Ising ring when the
ancillary spins are located at fixed distance $b$, therefore their
number scales as $n=L/b$.

Similarly to the CQT case, on the basis of the large-$n$ scaling
variable $Y_\kappa^\prime $ introduced in Eq.~\eqref{Ykappap}, we argue
that the appropriate scaling variable when $n=L/b$ is simply obtained
by replacing $n$ with $L/b$, thus
\begin{equation}
  \widetilde{Y}_\kappa = {k L^{1/2}\over b^{1/2} \Delta_{\cal I}(g,L)}
  \label{tildeYkappa}
\end{equation}
The other scaling variable $Y_\delta$ should be appropriate also in this
case.  Therefore the decoherence factor is expected to scale as
\begin{equation}
  Q(b,g,\kappa,\delta,L) \approx \widetilde{\cal
    Q}(\widetilde{Y}_\kappa,Y_\delta)\,.
\label{purityscafo3}
\end{equation}
Numerical results displayed in Fig.~\ref{figQscalingfirstorder2b}
provide evidence of this scaling behavior, within corrections
that appear to decay as $L^{-1}$.

\section{Behavior at the disordered phase}
\label{disbeh}

We finally discuss the effects of the interaction with the ancillary
system within the disordered phase of the Ising ring.

\begin{figure}[!t]
  \includegraphics[width=0.95\columnwidth]{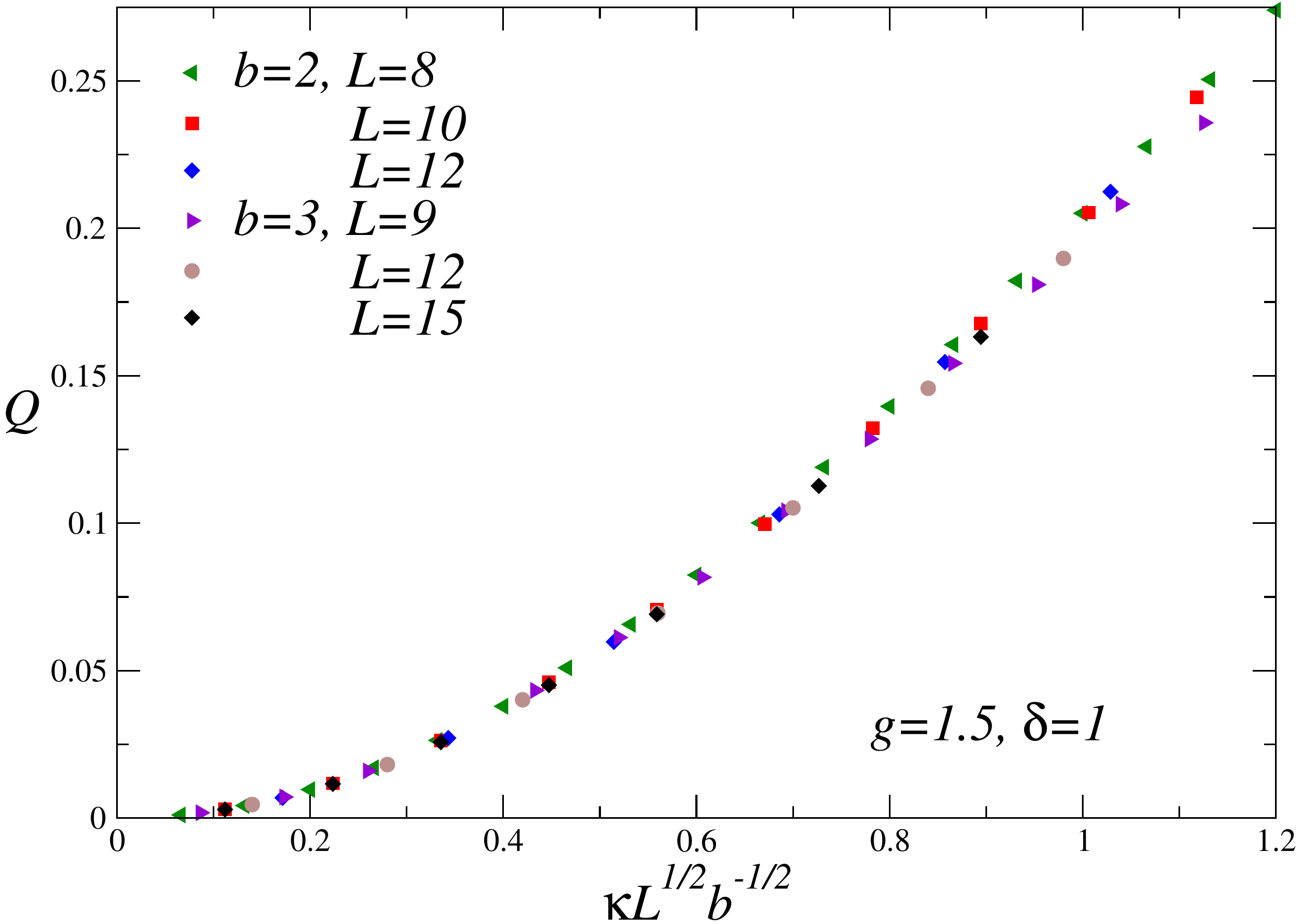}
  \caption{Decoherence function $Q$ versus $\kappa L^{1/2} b^{-1/2} =
    \kappa \sqrt{n}$ at the disordered phase, with $g=1.5$ and
    $\delta=1$. These data clearly support the $\sqrt{n}$-law of the
    dependence of the observables in the large-$n$ limit, and in
    particular of the decoherence factor.}
  \label{figQversuskparamagnetic}
\end{figure}

In this case we expect a trivial behavior when keeping $n$ fixed,
i.e., a substantial independence of the size $L$ when increasing it,
and a smooth behavior as a function of $\kappa$ and $\delta$. However,
it is interesting to note that, when increasing $n$, the
$\sqrt{n}$-law emerges also within the disordered phase. Indeed the
behavior is generally characterized by a dependence on
$\sqrt{n}\, \kappa$ for sufficiently large $n$, and therefore on
$\kappa L^{1/2}b^{-1/2}$ when we consider ancillary spins at fixed distance.
In Fig.~\ref{figQversuskparamagnetic} numerical results for $b=2$ and
$3$ support this hypothesis.

According to previous sections, a reasonable doubt might arise
regarding the probed phase of the lattice model, as the continuous
transition line moves to higher values $g_c>g_{\cal I}$ when $b$ is fixed,
cf.~Fig.~\ref{fig:phasediagram}.  However note that, with increasing
the lattice size, we also need to rescale $\kappa \sim L^{-1/2}$ to
maintain $\kappa L^{1/2}b^{-1/2}$ fixed. Thus, in the thermodynamic
limit, we only explore the phase diagram in the neighborhood of
$\kappa=0$, where the system is always disordered for any $g>g_{\cal I}$.

%%%%%%%%%%%%%%%%%%%%%%%%%%%%%%%%%%%%%%%%%%%%%%%%%%%%%%%%%%%%%%%%%%%%
%%%%%%%%%%%%%%%%%%%%%%%%%%%%%%%%%%%%%%%%%%%%%%%%%%%%%%%%%%%%%%%%%%%%

\section{Summary}
\label{conclu}

We have investigated the ground-state properties of a system of
quantum spin-$1/2$ chains arranged in a sunburst-like geometry, as
sketched in Fig.~\ref{setup}B), where the $L$ spins in the ring are
supposed to be described by a one-dimensional quantum Ising
model~\eqref{Iring}. The latter represents a prototypical quantum
many-body system showing different quantum phases, separated by FOQTs
and a CQT, when varying the intensity of the external transverse and
longitudinal fields.
Although the full system is considered to be isolated and thus
governed by a unitary evolution, its subparts (namely, the Ising ring
and the external ancillary isolated spins) are naturally subject to
decoherence, as unveiled by the decoherence factor $Q$~\eqref{Qdef}
and its susceptibility $\chi_Q$~\eqref{chiqdef}.

In this first exploratory study of the sunburst Ising model, we
address the properties of its subsystems within the ground state of
the global system, under different conditions controlled by the
various Hamiltonian parameters, as the Ising transverse field $g$, the
energy scale $\delta$ of the ancillary system, and the interaction
strength $\kappa$ between the two subsystems.  Even though we discuss
the behavior in the large-size limit, we essentially consider
finite-size systems exploiting FSS frameworks, thus the equilibrium
ground-state properties can be associated with the adiabatic dynamics.
Finite-size systems generally have a nonvanishing gap, therefore it is
always possible to conceive adiabatic evolutions for sufficiently
large time scales, even close to the quantum transitions.

Substantially different regimes emerge. At small $\kappa$ they are
related to the different phases of the Ising ring.  We observe rapid
changes with respect to variations of the Hamiltonian parameters close
to quantum transitions, such as along the FOQT line ($|g|<g_{\cal I}$) and at
the CQT ($g\simeq g_{\cal I}$) of the Ising ring, with the emergence of
peculiar scaling regimes. To perform this analysis we exploit RG and
FSS frameworks~\cite{RV-21}, which allow us to effectively describe
the behavior of systems in proximity of either CQTs or FOQTs.
We distinguish two notable large-$L$ limits where FSS laws develop:
(i) the case in which the number $n$ of isolated spins of the
ancillary system is kept fixed; (ii) the case in which their number
increases as $n=L/b$, being located at fixed spatial intervals of size
$b$, to maintain a residual translation invariance.  In the following
we summarize the main results.

For sufficiently small values of the Hamiltonian parameters $\delta$
and $\kappa$, the decoherence factor and also the correlations along
the Ising ring show nonanalytic behaviors at the CQT, controlled by
scaling variables associated with the Hamiltonian parameters $\kappa$
and $\delta$. The one associated with the interaction strength
$\kappa$ is given by (i) $K=\kappa L^{y_\kappa}$, where
$y_\kappa=7/8$, for a finite number of ancillary spins (the RG
exponent $y_\kappa$ is related to that of the symmetry-breaking
defects~\cite{FRV-22}), and (ii) $\widetilde{K}= \kappa
L^{\tilde{y}_\kappa}$ with $\tilde{y}_\kappa=y_\kappa+1/2=11/8$, for
an infinite number of ancillary spins.  The crossover between the two
regimes is essentially controlled by the peculiar $\sqrt{n}$
dependence when increasing $n$, which can be reabsorbed by an
appropriate redefinition of the scaling variable associated with the
Hamiltonian parameter $\kappa$, i.e. $K^\prime = \sqrt{n}
K$. Therefore, the scaling variable $\widetilde{K}$ reflects the fact
that the large-$n$ behavior is essentially controlled by the scaling
variable $\sqrt{n} K$ leading to $\widetilde{K}$ when replacing
$n=L/b$.  The scaling variable associated with the ancillary energy
scale $\delta$ turns out to be $A=\delta L^z$ with $z=1$, in both
cases (i) and (ii).  The scaling behavior of the decoherence factor,
reported in Eqs.~\eqref{puritysca} and~\eqref{qscanb}, implies rapid
changes when moving $\kappa$ and $\delta$ from zero, due to the
divergence of the corresponding susceptibility
[cf.~Eq.~\eqref{chiqdef}], as (i) $\chi_Q\sim L^{2y_\kappa}$ for
finite $n$, and (ii) $\chi_Q\sim L^{2\tilde{y}_\kappa}$ for $n=L/b$
(keeping $A$ fixed).

The impact on the decoherence behavior in the small-$\kappa$ and
$\delta$ regimes appears even more drastic at the FOQT transition
line, where the gap $\Delta_{\cal I}$ of the pure Ising ring gets suppressed
exponentially when increasing $L$.  The scaling variables turn out to
be (i) $Y_k= \kappa/\Delta_{\cal I}$ for finite $n$, and (ii)
$\widetilde{Y}_k= \kappa L^{1/2}/(b^{1/2}\Delta_{\cal I})$ for $n=L/b$, while
$Y_\delta=\delta/\Delta_{\cal I}$ in both cases (i) and (ii). As a
consequence, the decoherence susceptibility now diverges
exponentially, i.e., $\chi_Q\sim \Delta_{\cal I}^{-2}\sim e^{cL}$.

The above behaviors change when considering the large-size limit while
keeping the energy scale $\delta>0$ of the ancillary subsystem fixed.
In particular, we observe a smooth dependence of the various
quantities on $\kappa$ even around $\kappa=0$, when the number $n$ of
ancillary spins is finite. However, this is effectively controlled by
the product $\sqrt{n}\,\kappa$ when $n$ increases, and by $\kappa
L^{1/2}$, when $n$ increases linearly as $n=L/b$.  This leads to the
predictions that the decoherence susceptibility remain finite [$\chi_Q
  = O(1)$] at finite $n$, while it increases as $\chi_Q\sim L$ for
$n\sim L$.  Likewise we observe an analogous behavior of $\chi_Q$ with
$L$, at the disordered phase of the Ising ring (i.e., when $g>g_{\cal I}$).

Finally, we have analyzed the global phase diagram of the model in the
space of the Hamiltonian parameters $g,\,\kappa,\,\delta$.  As
expected from considerations based on the global symmetry of the
system, which maintains a ${\mathbb Z}_2$ symmetry,
cf. Eq.~\eqref{z2symm}, there is a line of Ising-like CQTs separating
disordered and ordered phases. This line runs for
$g_c(\kappa,\delta)>g_{\cal I}$, and in particular close to $\kappa=0$ it
behaves as $g_c(\kappa,\delta)\approx g_{\cal I} + C_\kappa \kappa^2$.
Around it, the decoherence factor behaves (i) as a smooth function of
$\kappa$ for finite $n$, or (ii) as $\kappa L^{1/2}$ for $n=L/b$.

Our analysis has focused on the sunburst Ising model with a
one-dimensional geometry (namely, the subsystem made of $L$
interacting spins is effectively a ring).  However, from a conceptual
point of view, it is straightforward to generalize such model to
higher dimensions (for example, the two-dimensional case corresponds
to an interacting spin system on a square lattice, where some spins
are coupled with isolated qubits, following a regular pattern).
Unfortunately the lack of integrability of such model drastically
limits the possibilities of standard numerical approaches up to very
small sizes.

It would be also tempting to investigate the sunburst model under
dynamic out-of-equilibrium protocols, for examples arising from sudden
quenches of one Hamiltonian parameter, as those already considered
within the more familiar central spin models (see, e.g.,
Ref.~\cite{RV-21}).  In this context, one could study peculiar
mechanisms arising in composite quantum systems, such as the onset of
decoherence in time, or the exchange of heat and work between the
various subportions of the whole system, characterizing their quantum
thermodynamic properties.

We point out that all the results obtained in this paper have been
carefully checked by means of numerical simulations for systems with a
limited amount of coupled qubits ($L \lesssim 20$).  This suggests the
possibility to devise near-term experiments with quantum simulators to
access our predictions directly in the lab, e.g., by means of trapped
ions~\cite{Islam-etal-11, Debnath-etal-16}, ultracold
atoms~\cite{Simon-etal-11, Labuhn-etal-16}, or superconducting
qubits~\cite{Salathe-etal-15, Cervera-18}.


\begin{thebibliography}{99}
\bibitem{corrQ1n2n} In actual computation we have used
  $\kappa=10^{-4}$ to evaluate $\chi_Q$. Systematic errors are
  estimated by comparing the discrepancies between data at
  $\kappa=10^{-3}$ and $10^{-4}$.
\end{thebibliography}
\end{document}